\documentclass[a4paper,fleqn]{cas-sc}
\usepackage[round,authoryear]{natbib}
\usepackage{graphicx}
\usepackage{booktabs}  
\usepackage{tabularx}
\usepackage{array}
\usepackage{amsmath}
\usepackage[ruled]{algorithm2e}
\usepackage{algorithmic}
\usepackage{cleveref}
\usepackage{threeparttable}
\usepackage{multirow}
\newcommand{\myref}[1]{(\ref{#1})}
\crefname{section}{Section}{Sections}
\Crefname{section}{Section}{Sections}

\SetAlFnt{\normalfont\small}
\SetAlCapFnt{\normalfont\small}
\SetAlCapNameFnt{\normalfont\small}
\SetNlSty{textnormal}{}{}
\ExplSyntaxOn
\cs_set:Npn \__make_tbl_caption:nn #1#2
{
  \l_tbl_align_tl
  \skip_vertical:N \l_tbl_abovecap_skip
  {\parbox{\dimexpr(\l_tbl_width_dim)}
    {\rightskip=0pt\normalfont\small\textbf{\color{scolor}#1}\par#2\par\vskip4pt}}
  \skip_vertical:N \l_tbl_belowcap_skip
}
\cs_set:Npn \__make_fig_caption:nn #1#2
{
  \l_fig_align_tl
  \skip_vertical:N \l_fig_abovecap_skip
  \setbox\cascaptionbox=\hbox{%
     \normalfont\small\textbf{\color{scolor}#1:}~#2}
  \ifdim\the\wd\cascaptionbox<\dim_use:N \l_fig_width_dim\relax
    \parbox{\l_fig_width_dim}
      {\unskip\ignorespaces\hfil\normalfont\small
       \textbf{\color{scolor}#1:}~#2\hfil\par}
  \else
    \parbox{\l_fig_width_dim}
      {\rightskip=0pt\unskip\ignorespaces\normalfont\small
       \textbf{\color{scolor}#1:}~#2\par}
  \fi
  \skip_vertical:N \l_fig_belowcap_skip
}
\cs_set:Npn \__first_footerline:
{
  \group_begin:
  \normalfont\small
  \ifnum\theblind>0\relax
  \else
    \__short_authors: :~
  \fi
  { \itshape Preprint~ submitted ~to ~Elsevier }
  \group_end:
}
\cs_set:Npn \__cas_head:
{
  \parbox{\textwidth}
    {
      \normalfont\small\centering
      \__short_title:
    }
}
\cs_set:Npn \__cas_foot:
{
  \parbox[t]{\textwidth}
  {
   \rule{\textwidth}{.2pt}\\
   \normalfont\small
   \__first_footerline:
   \hfill Page~\thepage {}~of~ \lastpage
  }
}
\ExplSyntaxOff

\begin{document}
	\let\WriteBookmarks\relax
	\shorttitle{Non-invasive Blood Glucose Estimation from Wearable Physiological Signals}
	\shortauthors{Anonymous Authors}
	\shortauthors{Zhang et~al.}

	\author[1]{Zexing Zhang}[orcid=0009-0001-9938-4766]
	
	\author[1]{Jichao Li}[orcid=0009-0009-2087-434X]

	\author[2,3]{Yilong Wang}
	
	\author[1]{Kewei Yang}[orcid=0000-0001-7090-9146]
	
	\author[1]{Tianyang Lei}[orcid=0000-0002-4405-919X]	\cormark[1]
	\ead{leitianyang20@163.com}
	\cortext[1]{Corresponding author.}
	\affiliation[1]{organization={College of Systems Engineering},
		addressline={National University of Defense Technology}, 
		city={Changsha},
		postcode={410000}, 
		country={China}}
	\affiliation[2]{
		addressline={The First Hospital of Jilin University}, 
		city={Jilin},
		postcode={130021}, 
		country={China}}
	\affiliation[3]{organization={School of Mathematics and Statistics}, 
		addressline={Changchun University of Technology},
		city={Jilin},
		postcode={130102}, 
		country={China}}
	\title [mode = title]{Dynamic Incremental Learning for Non-invasive Blood Glucose Estimation from Wearable Physiological Signals}
	\begin{abstract}
		Non-invasive blood glucose estimation from wearable physiological signals remains difficult because longitudinal photoplethysmography (PPG) data are subject to distribution drift, whereas reference capillary blood glucose labels are sparse and costly to acquire. We propose a deep-learning-based dynamic incremental learning (DIL) framework that combines a mutual entropy-optimized replay-based dynamic clustering module (MERDC) with an uncertainty-quantified proxy gradient bridging agent (PGBA) for label-efficient adaptation to unlabeled PPG streams. To support this setting, we further establish a longitudinal benchmark dataset comprising PPG, reference capillary blood glucose, and cuff blood pressure measurements from 183 participants collected over 285 days, and we make this resource available to the research community. Under 5-fold subject-independent validation, the proposed method achieves a mean absolute error (MAE) of $0.64 \pm 0.01$ millimoles per liter (mmol/L) and a root mean square error (RMSE) of $1.29 \pm 0.10$ mmol/L, with $97.69 \pm 1.63\%$ of estimates falling within Clarke zones A+B. Aggregation-level analyses further support the robustness of the observed error distribution beyond window-level evaluation. These results provide a proof-of-concept for adaptive non-invasive glucose estimation in wearable physiological sensing and establish a longitudinal benchmark for subsequent research.
	\end{abstract}

	\maketitle

	\section{Introduction}
	\label{Introduction}
	Diabetes represents a major global public health challenge and remains a leading cause of mortality and disability. It is projected that more than 1.31 billion people worldwide will be living with diabetes by 2050 \citep{c:1}. Conventional blood glucose measurement methods, including fingerstick blood sampling and continuous glucose monitoring (CGM), are invasive, causing both psychological and physical discomfort to patients. In addition, these methods require frequent replacement of consumables, resulting in high costs \citep{c:2}. Therefore, the development of a long-term, accessible, and minimally painful blood glucose estimation approach remains an important research objective. Such an approach could broaden access to longitudinal glucose assessment and support earlier risk screening in lower-burden settings.
	
	Diabetes is associated with vascular endothelial dysfunction and autonomic neuropathy, which result in abnormal vasomotor regulation, increased arterial stiffness, and reduced pulse rate variability. Among wearable physiological sensing modalities, non-invasive photoplethysmography (PPG) signals can capture these physiological alterations \citep{c:3,c:5,c:8,c:9}. A PPG signal is generated by light absorption measured through an optical sensor placed on the skin, typically at the fingertip or wrist. It comprises two primary components: a pulsatile component, which correlates with the cardiac rhythm, and a non-pulsatile component. The pulsatile component provides information on cardiovascular blood flow \citep{c:4}, whereas the non-pulsatile component reflects baseline blood volume, sympathetic nervous system activity, and vasomotor regulation \citep{c:5}. 
	
	Moreover, because PPG acquisition devices are low-cost and portable, they provide an attractive sensing modality for less invasive glucose estimation research. Nevertheless, the complex and nonlinear relationship between PPG signals and blood glucose levels remains a major challenge. Accordingly, leveraging the powerful feature representation capabilities of deep learning (DL) models to uncover the implicit correlations between PPG signals and blood glucose is a promising research direction \citep{c:6,c:7,c:8,c:9,c:10,c:11,c:12,c:13}.
	However, applying DL models to PPG signals for individual blood glucose estimation presents three major challenges.
	First, intra-individual physiological regulation introduces persistent non-stationarity. Dietary intake, physical activity, and pharmacological interventions continuously reshape the distributional and temporal characteristics of blood glucose and PPG signals \citep{c:6,c:8,c:17}. Conventional approaches are poorly suited to accommodate such dynamics without complete model retraining on static datasets, a step that is both computationally and memory prohibitive \citep{c:22}. Although incremental learning (IL) provides a theoretical pathway to alleviate this burden, mainstream IL frameworks were originally designed for class- or task-incremental classification with explicit task boundaries and dense supervision \citep{c:20}. These frameworks typically assume discrete semantic accretion, quasi-stationary intra-task distributions, and enumerable replay scheduling. Such assumptions are incompatible with our setting. Drift is continuous, multi-scale, and regression-oriented rather than proceeding via stepwise class expansion. The latent tasks, understood here as physiological regimes, are not pre-specified and must be discovered adaptively. Labeled glucose samples are moreover scarce and costly, whereas unlabeled PPG windows arrive in continuous streams. Moreover, naïve adaptive schemes such as periodic full retraining, sliding-window replacement, or unstructured replay either erode long-term physiological memory, amplify transient perturbations, inflate computational overhead, disregard unlabeled modalities, or lack principled stability–plasticity regulation \citep{c:61,c:62,c:63}.
	
	To address these physiological and methodological limitations, we propose Dynamic Incremental Learning (DIL), a domain-specialized paradigm that integrates mutual-entropy-guided latent task discovery, bounded episodic replay with gradient non-interference projection, and uncertainty-aware proxy-gradient exploitation of unlabeled PPG signals. This framework enables continual, label-efficient, and computation-aware adaptation while preserving long-term physiological memory and predictive stability, thereby offering the potential for personalized and robust blood glucose estimation in longitudinal health monitoring.
	
	Second, existing blood glucose estimation methods assume a fixed data distribution, rendering them inadequate for handling uncertainties arising from dynamic variations across individuals, environments, and devices \citep{c:2,c:7,c:10}. Such variations include differences in lifestyle, climate, PPG acquisition hardware, and unavoidable random measurement errors in reference physiological parameters \citep{c:6,c:7,c:9}. We conceptualize these dynamics as a sequence of heterogeneous tasks with unknown categories. Beyond meta-task clusters, newly arriving data may also form derived-task clusters, which refine existing clusters when the model encounters novel signal patterns. This setting fundamentally differs from mainstream incremental learning, which presumes predefined task categories and a corresponding partitioning \citep{c:20}.
	
	To address this partitioning challenge, we design a workflow that optimally adapts to new data while preserving the integrity of existing clusters. At its core is a mutual entropy-optimized replay-based dynamic clustering method that employs a unified objective to balance intra-cluster signal entropy and inter-cluster mutual information. In doing so, it effectively segments meta-task clusters and deploys them within a replay-based dynamic strategy, thereby maximizing the benefits of incremental learning with minimal computational overhead.
	
	Third, a fundamental mismatch exists between invasively obtained blood glucose labels and readily acquired, unlabeled PPG signals. This disparity yields a vast volume of PPG data without corresponding blood glucose annotations \citep{c:ppgpt,c:15,c:11,c:23}. Acquiring large-scale, high-quality labeled data in this domain is prohibitively expensive and time-consuming \citep{c:18,c:16,c:13}. To operate under such resource constraints, we propose a Proxy Gradient Bridging Agent (PGBA) grounded in uncertainty quantification. PGBA first trains on a small labeled set and then incrementally incorporates newly arriving unlabeled and labeled data. This process reduces reliance on large labeled datasets and enhances the model’s generalization. PGBA introduces a bridging-agent mechanism: it learns uncertainty-quantified blood glucose gradients from limited labeled data and then predicts proxy gradients for unlabeled data, thereby integrating the unlabeled data into the incremental learning process.
	In summary, we establish a new paradigm for non-invasive blood glucose estimation using PPG signals, specifically designed for real-world scenarios. This paradigm integrates a newly established longitudinal benchmark dataset, a novel method for data-distribution partitioning, and a dynamic incremental learning framework, all tailored to the challenges of repeated physiological measurement under realistic conditions. Our main contributions are:

	\begin{enumerate}[(1)]
		\item We introduce dynamic incremental learning to this field for the first time. This approach leverages episodic memory and gradient projection to learn a common, relevant subset from dynamically distributed data. It adapts to data drift driven by intra-individual physiological regulation and to uncertainties arising from environmental or device variations, establishing a new pathway for longitudinal physiological monitoring.
		\item We design a mutual-entropy-optimized, replay-based dynamic clustering method. This method balances intra-cluster signal entropy with inter-cluster mutual information to partition the meta-task and derived-task clusters required for incremental learning. This process supports optimal adaptability to dynamic data distributions in real-world blood glucose estimation.
		\item We propose a Proxy Gradient Bridging Agent (PGBA) that employs uncertainty quantification to enable label-efficient incremental learning with unlabeled PPG signals. PGBA reduces the costs and labor associated with invasive blood glucose measurements and manual data labeling.
		\item We establish and release a longitudinal benchmark dataset for this task. The dataset includes PPG signals, reference capillary blood glucose values, and cuff blood pressure measurements from 183 participants collected over 285 days, spanning participants with a mean age of $57.02 \pm 21.20$ years and a clinically relevant glucose range of 4.2--16.3 mmol/L.
		\enlargethispage{-\baselineskip}
	\end{enumerate}
	
	The remainder of this paper is organized as follows. Section~\ref{sec:related work} reviews related work on physiological measurements for blood glucose estimation and incremental learning paradigms. Section~\ref{sec:methods} describes the methods and materials, including the benchmark dataset construction, the proposed dynamic incremental learning framework, the mutual entropy-optimized replay-based dynamic clustering method, and the proxy gradient bridging agent. Section~\ref{sec:experiments} presents the experimental results, including implementation details, evaluation criteria, comparisons with state-of-the-art methods, and ablation studies. Section~\ref{sec:discussions} discusses long-term validation, translational relevance, catastrophic forgetting, interpretability, robustness in real-world scenarios, assumptions, and broader implications. Finally, Section~\ref{sec:conclusions} concludes the paper and outlines future directions.
	\section{Related Work}
	\label{sec:related work}
	\subsection{Physiological Measurement for Blood Glucose}
	Most studies on blood glucose monitoring using PPG signals have primarily focused on specific glucose ranges \citep{c:10,c:38,c:19,c:40}, such as classifying risks of normal, hypoglycemia, and hyperglycemia. However, these methods do not provide comprehensive, round-the-clock information on glucose variability, which is critical for understanding diabetes-related complications \citep{c:8,c:43}. With the growing demand for precision medicine and dynamic physiological monitoring, methods for real-time blood glucose estimation have gradually become a focal point. These methods offer greater immediacy and can improve treatment adaptability.
	
	Current research mainly encompasses traditional machine learning \citep{c:24,c:25,c:10,c:21,c:11,c:18}, deep learning \citep{c:6,c:7,c:19,c:37,c:38,c:36}, multi-model fusion \citep{c:8,c:9,c:13}, and personalized modeling \citep{c:23,c:33} approaches. Traditional machine learning methods rely on handcrafted features extracted from raw PPG signals, including time-domain features such as pulse transit time \citep{c:11}, DC component level \citep{c:25}, and rise time \citep{c:33}, as well as statistical features such as power spectral entropy \citep{c:8}, peak amplitude \citep{c:10}, and root mean square \citep{c:12}. While these methods offer a degree of interpretability, simple feature analysis is insufficient to capture the underlying relationship between PPG signals and blood glucose levels effectively \citep{c:9}. Moreover, the limited scope and generalizability of traditional features constrain their ability to represent inter- and intra-individual variability.
	
	In recent years, deep learning and multi-model fusion methods have emerged as effective solutions to address limited representational capacity. \citet{c:36} developed a CNN-based central state-aware algorithm that enables non-invasive continuous blood glucose monitoring without handcrafted features. However, this method overlooks the aggregation of adaptive contextual information. To address this, \citet{c:9} designed a convolutional multi-view attention block and explored a broader search space and dependencies through a cascaded bidirectional long short-term memory network. Given the inherently nonlinear and non-stationary nature of PPG signals, \citet{c:7} proposed a feature enhancement module to improve the network’s discriminative ability for periodic and trend patterns. To further identify implicit physiological patterns, \citet{c:6} reconstructed and embedded kinematic features based on spatial position encoding as prior knowledge, while \citet{c:13} used causal network analysis to fuse higher-order information, encouraging the decomposition of latent physiological factors into endogenous associations aligned with downstream fusion information. \citet{c:8} adopted a multi-model approach based on weighted Choquet integrals and spatiotemporal feature fusion, combining the feature extraction capabilities of traditional machine learning and deep learning. However, this method requires additional ECG signals, which are difficult to collect in portable settings. \citet{c:23} introduced a deductive learning-based blood glucose monitoring method on limited training datasets, exploring personalized physiological measurement modeling. \citet{c:15} proposed TS2TC, a self-supervised pretraining framework that improves blood glucose estimation accuracy from PPG under limited labeled data by learning generalizable physiological representations. Building on large-scale PPG representation learning, \citet{c:ppgpt} introduced PPGPT, which transfers next-token language modeling to PPG signals. Both methods, however, assume a static training distribution and do not address the longitudinal distribution drift and label scarcity that characterize real-world continuous monitoring. Bridging this gap is the central motivation for the present work.
	
	In summary, data drift driven by intra-individual physiological regulation and dynamic uncertainties arising from variations across individuals, environments, and devices remain two critical challenges that require urgent attention. Additionally, many studies rely on small, limited datasets. For example, in deep learning studies, the arithmetic mean and standard deviation of participant counts are approximately $54 \pm 91$ \citep{c:7,c:37,c:9,c:6,c:12,c:8,c:33}. By contrast, our study establishes a longitudinal benchmark spanning 183 participants, accompanied by comprehensive experimental analyses.
	\subsection{Incremental Learning}
	Humans and other animals can gradually learn new skills without compromising previously acquired ones \citep{c:49}. In stark contrast, despite the remarkable achievements of deep models in many closed-world tasks, such as CNNs \citep{c:47} and Transformers \citep{c:48}, these models often rapidly lose previously acquired capabilities when trained on new tasks or data distributions, failing to accurately approximate the optimal expected risk minimization function. These two phenomena, namely catastrophic forgetting and overfitting, stem from non-stationary data streams in the real world \citep{c:44,c:46}. They correspond to two core challenges in incremental learning: the stability–plasticity dilemma and unreliable empirical risk minimization. Incremental learning (IL) aims to bridge the capability gap between natural and artificial intelligence \citep{c:44}.
	
	The current mainstream paradigm in incremental learning is Class-Incremental Learning (CIL). CIL represents a more challenging scenario in open-world settings \citep{c:20}, where training data arrive sequentially in a streaming format. At each timestamp, a new training dataset (a ``task'') is obtained, and the model must be updated with new classes. However, this paradigm does not effectively capture the temporal dynamics of underlying data distributions in physiological measurement scenarios. We therefore reshape a paradigm tailored to physiological measurement contexts. It is also worth noting that an ideal IL paradigm should be compatible with any backbone network \citep{c:20}, including CNN-based \citep{c:8,c:36} and Transformer-based \citep{c:9,c:6} methods already proposed for blood glucose estimation.
	\begin{figure}[pos=htbp]
		\centering
		\includegraphics[width=0.95\linewidth]{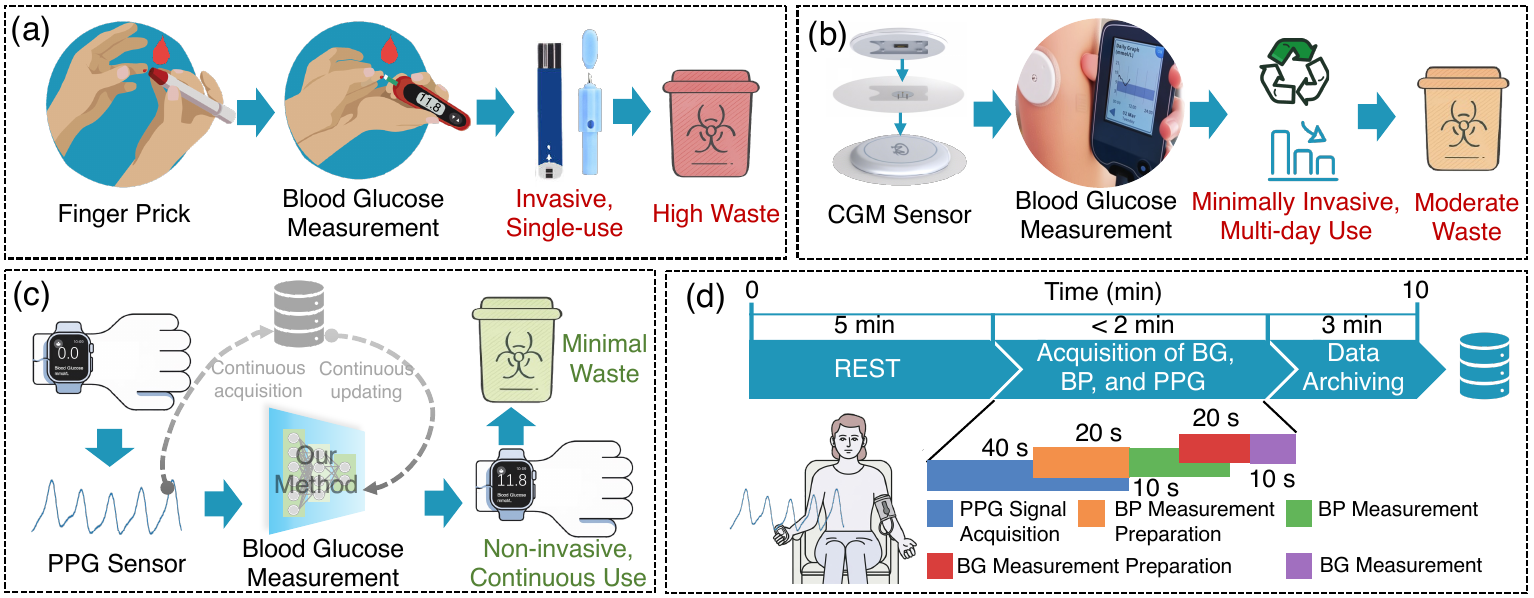}
		\caption{Illustrating the technological progression from (a) traditional invasive fingerstick sampling and (b) Continuous Glucose Monitoring (CGM) to (c) our proposed non-invasive method. The feasibility of this approach is established through (d) our systematic data collection protocol, which produced a novel public dataset.}
		\label{fig1}
	\end{figure}
	\section{Methods and Materials}
	\label{sec:methods}
	Fig. \ref{fig1} illustrates the progression from invasive, single-use fingerstick measurement to minimally invasive, multi-day CGM in panels (a) and (b), the proposed fully non-invasive, high-frequency PPG-based approach that eliminates puncture and enzyme strips in panel (c), and the structured data acquisition protocol integrating synchronized PPG signals, reference capillary blood glucose, and blood pressure measurements in panel (d). This continuum provides the longitudinal, distribution-shifting signal environment required to address the challenges outlined in Section \ref{Introduction}.
	
	To address the three challenges from Section \ref{Introduction}, we propose an integrated framework that estimates blood glucose non-invasively from PPG signals. This framework combines dynamic incremental learning with an uncertainty-aware approach to leveraging unlabeled data. The first two challenges, namely data drift due to physiological changes and the limitations of fixed data distributions, create a stability–plasticity dilemma. This dilemma forces conventional models to undergo costly retraining for new data, risking catastrophic forgetting of existing knowledge. To mitigate this, we introduce a Mutual Entropy-Optimized Replay-Based Dynamic Clustering module (MERDC) for task discovery and replay-based adaptation under evolving data distributions. Within MERDC, the internal clustering operator Mutual Entropy-optimized Enhanced Adaptive Clustering (ME2AC) partitions streaming PPG–glucose data into distinct clusters via mutual entropy optimization, while replay-based updating preserves prior knowledge during adaptation to newly arriving patterns.
	
	The third challenge is the scarcity of labeled data, which leads to underutilization of abundant unlabeled PPG signals and limits reliable empirical risk minimization. To address this, we introduce a Proxy Gradient Bridging Agent (PGBA). PGBA learns uncertainty-quantified gradients from a small labeled dataset and predicts proxy gradients for unlabeled signals. It then integrates these proxy gradients into the incremental updates, enhancing model generalization with minimal labeling cost.
	
		\begin{figure}[pos=htbp]
		\centering
		\includegraphics[width=0.96\linewidth]{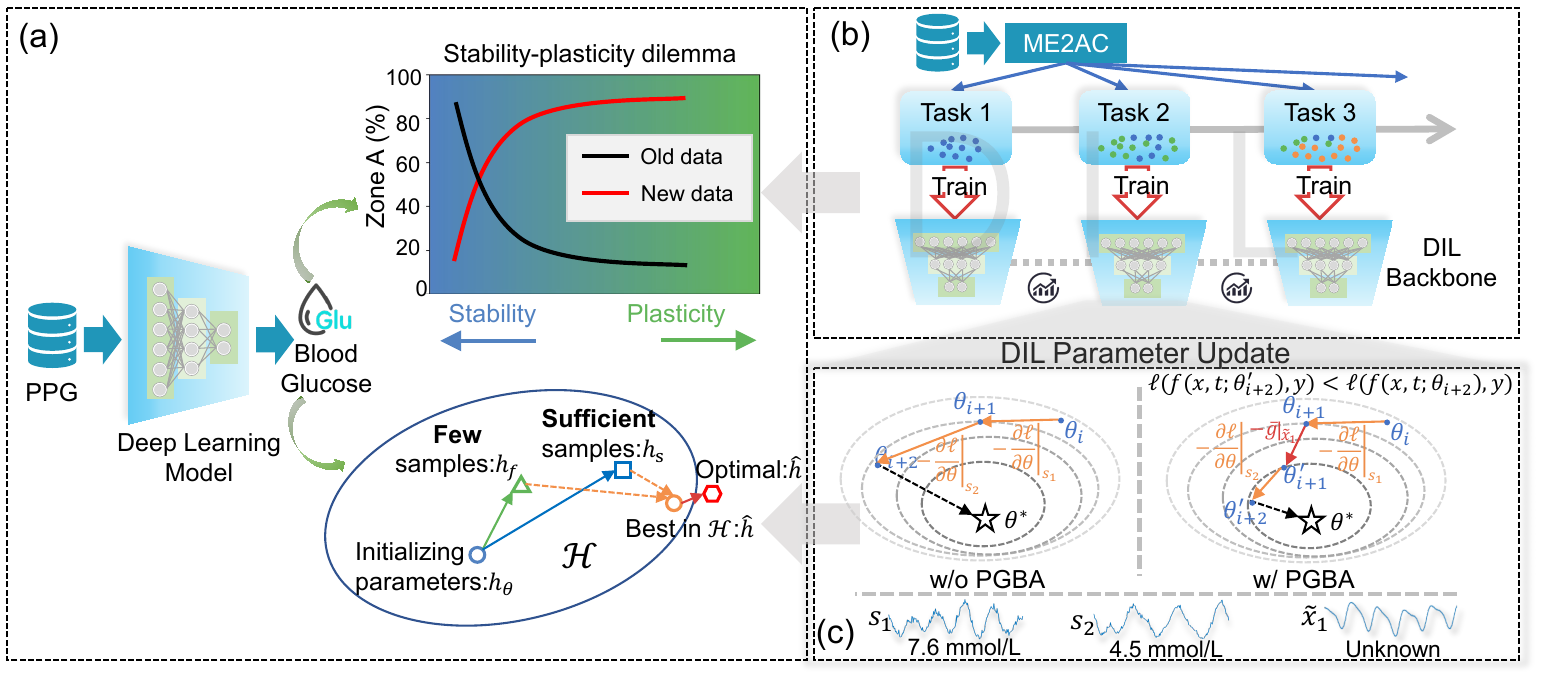}
		\caption{Overview of the proposed MERDC-enhanced dynamic incremental learning (DIL) framework for non-invasive blood glucose estimation from PPG signals. (a) Stability--plasticity dilemma and unreliable empirical risk minimization under limited labeled data. In the inset, the black curve represents old-data retention associated with stability, whereas the red curve represents new-data adaptation associated with plasticity. (b) The MERDC module, instantiated through the ME2AC clustering operator, mitigates catastrophic forgetting by organizing meta- and derived-task streams for replay-based dynamic learning. (c) Proxy Gradient Bridging Agent (PGBA) supports label-efficient empirical risk minimization by incorporating unlabeled PPG windows through uncertainty-quantified proxy gradients.}
		\label{fig2}
	\end{figure}
	
	Table \ref{tab:1} and Fig. \ref{fig9} present the statistical information of the dataset, while Fig. \ref{fig10} illustrates the dataset collection standards. Fig. \ref{fig2} presents a conceptual overview of the proposed framework. Panel (a) highlights the core challenges: the stability–plasticity dilemma arising from data drift and unreliable risk minimization due to limited labeled data. Panel (b) illustrates how the MERDC-enhanced dynamic incremental learning approach addresses the stability–plasticity dilemma, while panel (c) demonstrates how the PGBA module mitigates data scarcity by leveraging unlabeled PPG signals. To enable evaluation under realistic conditions, we also constructed a longitudinal benchmark dataset comprising PPG signals, reference capillary blood glucose measurements, and cuff-based blood pressure recordings from 183 participants over 285 days, spanning a broad demographic and clinical spectrum. In the following, we first describe the dataset collection and construction process.
	\begin{table}[pos=htbp] \centering \normalfont\footnotesize \setlength{\tabcolsep}{4pt} \caption{Summary of main abbreviations and symbols used in the manuscript.} \label{tab:abbr_symbol} \begin{tabularx}{\linewidth}{@{}lX@{}} \toprule Term & Meaning \\ \midrule PPG & Photoplethysmography signal used as the wearable physiological input. \\ BG & Reference capillary blood glucose concentration, reported in mmol/L. \\ DIL & Dynamic Incremental Learning, the proposed continual regression framework. \\ MERDC & Mutual Entropy-Optimized Replay-Based Dynamic Clustering module for task discovery and replay-based updating. \\ ME2AC & Mutual Entropy-Optimized Enhanced Adaptive Clustering, the clustering operator used inside MERDC. \\ PGBA & Proxy Gradient Bridging Agent, the module that learns bounded proxy gradients for unlabeled PPG windows. \\ CEG & Clarke Error Grid, used to assess clinical error-region distribution. \\ NLL & Negative log-likelihood used for probabilistic uncertainty-aware regression. \\ PICP & Prediction interval coverage probability for uncertainty calibration. \\ CI & Confidence interval. \\ $\mathbf{x}$ & Input PPG window. \\ $y$ & Continuous reference capillary BG label. \\ $\hat{y}$ & Predicted BG value. \\ $\hat{\sigma}^{2}$ & Predicted input-dependent variance. \\ $\phi(\cdot)$ & Backbone feature encoder. \\ $\mathcal{T}$, $B$ & ME2AC-discovered task set and the corresponding number of latent tasks. \\ $\mathcal{E}$, $\mathcal{M}_b$ & Episodic memory set and the task-wise memory subset for task $b$. \\ $C_{\mathrm{mem}}$ & Total replay memory capacity. \\ $\tau_H$, $\tau_{\mathrm{MI}}$ & Entropy and mutual-information thresholds used in ME2AC. \\ $S_{\mathrm{SQI}}$ & Skewness-based signal quality index used during waveform quality evaluation. \\ \bottomrule \end{tabularx} \end{table}
	
	\subsection{Benchmark Dataset}
The initial candidate cohort comprised 195 participants before quality-control screening. After the pre-specified curation procedure described in Fig.~\ref{fig10}, 183 participants were retained in the final longitudinal benchmark dataset, including 112 males and 71 females. Data collection occurred over a 285-day period from March 2024 to January 2025. The dataset contains synchronized PPG signals, reference capillary blood glucose measurements, and cuff-based blood pressure recordings, thereby supporting research into the relationship among endocrine status, cardiovascular dynamics, and wearable PPG morphology. Its scale and temporal span make it suitable for studying non-invasive glucose estimation and cuffless blood pressure analysis under repeated-measurement settings. To our knowledge, it is among the earliest longitudinal fingertip-PPG benchmarks of this scale in this field. Table~\ref{tab:1} provides a detailed statistical overview of participants' demographic data and physiological indicators. The study was reviewed and approved by the institutional ethics committee of the participating clinical site, with the institution name and approval identifier withheld for double-anonymized review. All participants provided informed consent before any data collection.
	
\begin{figure}[pos=htbp]
		\centering
		\includegraphics[width=1\linewidth]{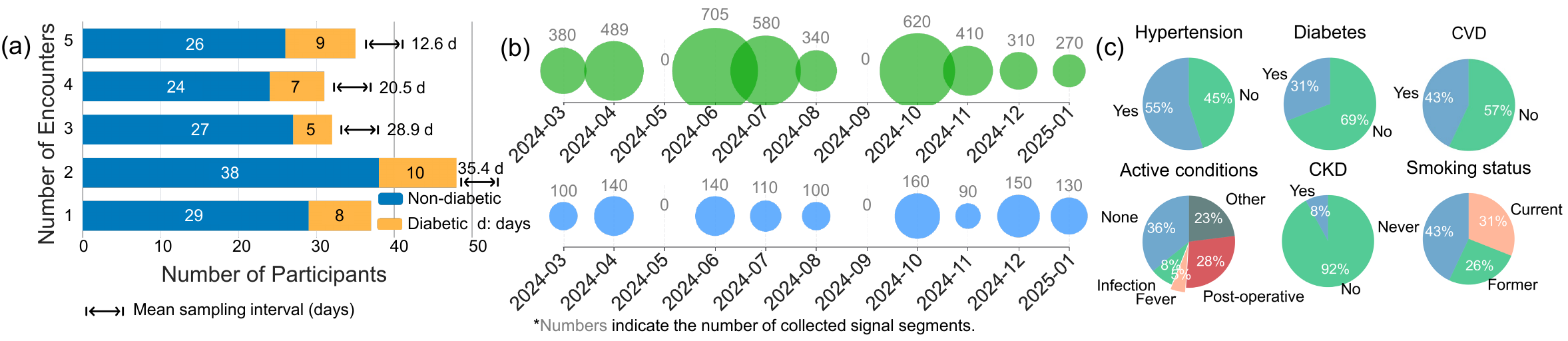} 
		\caption{Summary of encounter frequency, temporal distribution, and baseline health profiles in the longitudinal benchmark dataset. (a) Participant distribution by encounter count, stratified by diabetic and non-diabetic status. For groups with repeated encounters, the horizontal interval marker indicates the mean time interval between consecutive encounters, reported in days. (b) Monthly distribution of collected PPG signal segments from March 2024 to January 2025, showing seasonal variation and periods of intensive acquisition. (c) Baseline health and comorbidity characteristics at enrollment, including hypertension, diabetes, cardiovascular disease (CVD), chronic kidney disease (CKD), active clinical conditions, and smoking status. These metadata provide context for interpreting PPG signal variability under clinically realistic acquisition conditions.}
		\label{fig9}
	\end{figure}

\begin{figure}[pos=htbp]
		\centering
		\includegraphics[width=0.98\linewidth]{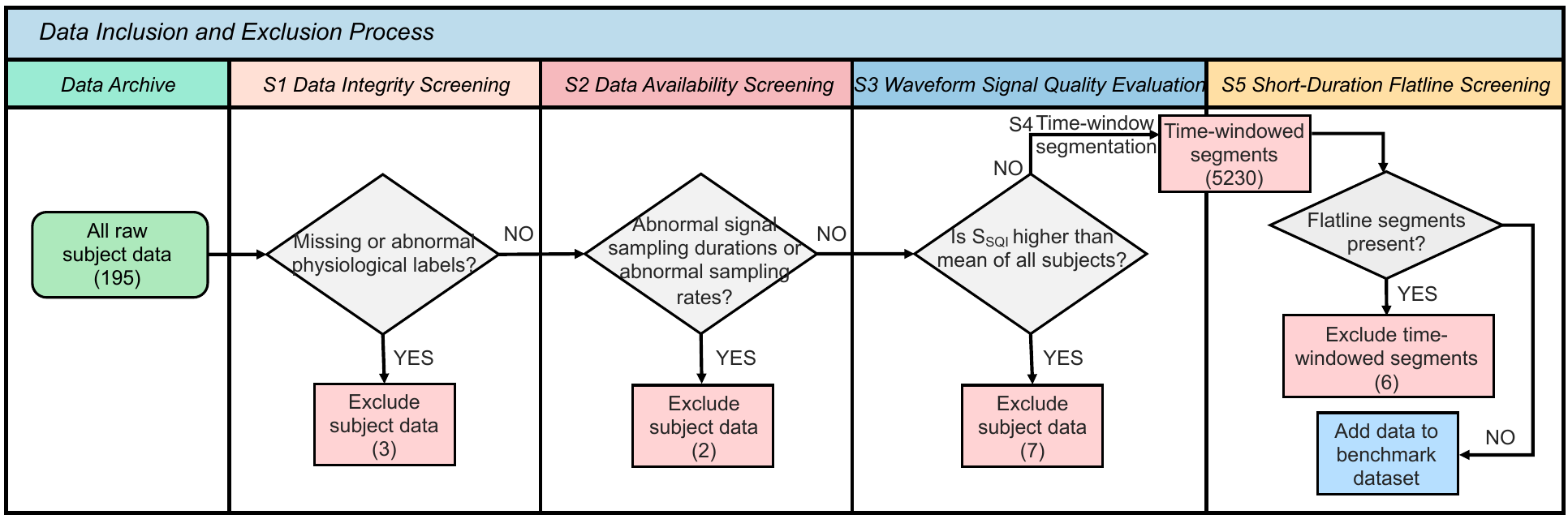}
		\caption{Curation pipeline for the study cohort ($n=195$): S1 Data integrity screening removed 3 subjects with invalid (out-of-range) or log-disconfirmed blood glucose or blood pressure labels. S2 Data availability screening removed 2 subjects whose raw 40\,s PPG segments violated either the 1000\,Hz~±0.5\% sampling rate or 40.0\,s~±1\% duration criteria. S3 Waveform signal quality evaluation computed $S_{\text{SQI}}$ on raw 40\,s segments and excluded 7 subjects with $S_{\text{SQI}}$ below the cohort mean. S4 Time-window segmentation retained data from 183 subjects, yielding 5,230 non-overlapping 4\,s windows. S5 Short-duration flatline screening removed 6 windows with variance $<10^{-5}$ for $\geq$0.5\,s. Final dataset: 183 subjects and 5,224 windows, retaining challenging but physiologically plausible morphologies.}
		\label{fig10}
	\end{figure}

	The dataset captures both inter-participant and intra-participant variability in encounter frequency and sampling density, as illustrated in Fig.~\ref{fig9}(a). Participants had between one and five encounters, and repeated-encounter groups showed different mean intervals between consecutive visits. For instance, participants completing five encounters had a mean interval of approximately $12.6$ days between visits, supporting the evaluation of short- and medium-term physiological variability. Fig.~\ref{fig9}(b) depicts the temporal distribution of collected PPG signal segments across the study period, highlighting seasonal fluctuations and periods of intensive data acquisition. Overall, the dataset contains encounters spaced from days to several months apart, enabling analyses across both short- and long-term physiological trends.
	Baseline health and comorbidity profiles are summarized in Fig.~\ref{fig9}(c). At enrollment, $55\%$ of participants presented with hypertension, $31\%$ with diabetes, and $43\%$ with cardiovascular disease, while chronic kidney disease (CKD) was reported in $8\%$ of participants. Active conditions during encounters included infections, fever, postoperative states, and other acute or chronic illnesses. Smoking status distribution comprised $43\%$ never-smokers, $26\%$ former smokers, and $31\%$ current smokers. These metadata provide essential context for interpreting PPG signal variability, as cardiovascular status, renal function, metabolic disorders, and acute illness can influence waveform morphology and derived indices.
	
	By combining encounter-level temporal resolution with detailed baseline health characterization, the dataset offers a rich foundation for robust algorithm development and evaluation under clinically realistic conditions.
	
	\begin{table}[pos=htbp]  
		\centering
		\normalfont\footnotesize
		\setlength{\tabcolsep}{1.6pt}
		\caption{Demographic and physiological statistics of the longitudinal benchmark dataset used in this study.}  
		\begin{threeparttable}
			\begin{tabular}{@{}lcccccccc@{}}  
				\toprule  
				Metric & Weight (kg) & Height (m) & Age (years) & BMI (kg/m\textsuperscript{2}) & HR (bp/m) & SBP (mm Hg) & DBP (mm Hg) & BG (mmol/L) \\
				\midrule  
				Mean & 69.76 & 1.69 & 57.02 & 24.35 & 72.98 & 130.02 & 71.74 & 6.56 \\
				Minimum & 46.30 & 1.55 & 21 & 16.80 & 55 & 85 & 44 & 4.2 \\
				Maximum & 98 & 1.87 & 96 & 33.51 & 102 & 164 & 94 & 16.3 \\
				Standard Deviation & 10.93 & 0.07 & 21.20 & 3.20 & 7.67 & 18.57 & 9.85 & 2.12 \\
				\bottomrule  
			\end{tabular}  
			{\normalfont\footnotesize BMI: Body mass index, HR: Heart rate, SBP: Systolic blood pressure, DBP: Diastolic blood pressure, BG: Blood glucose.}
		\end{threeparttable}
		\label{tab:1}  
	\end{table}
	
	\subsubsection{Experimental Design and Data Acquisition}
	The experimental protocol took place in an open environment. Participants sat in a relaxed state with their arms resting on a table. No constraints, such as medication use or fasting, were imposed, allowing the data to reflect dynamic intra-individual drift and external variability.
	
	Each data collection session lasted 10 minutes. Upon arrival, each participant had a five-minute acclimatization period to adjust to the environment and stabilize breathing. The acquisition setup was as follows: reference capillary blood glucose and PPG signals were collected from the tip of the left index finger, while arterial blood pressure was measured on the right forearm. All measurements were completed within a two-minute window. Fig. \ref{fig1}(d) illustrates this collection process. Specifically, reference capillary blood glucose was measured using the FDA-cleared Accu-Chek Performa system (Roche, Basel, Switzerland), which has a testing range of 0.6 to 33.3 mmol/L. The same device was used throughout the study, with daily L1/L2 quality-control verification after powering on. Arterial blood pressure was measured with the Omron J751 upper-arm monitor (Omron, Kyoto, Japan). PPG signals were recorded at a sampling frequency of 1 kHz using an EVAL-ADPD4100Z-PPG evaluation board and an EVAL-ADPDUCZ microcontroller (Analog Devices, San Jose, CA, USA).
	\subsubsection{Data Inclusion and Exclusion Criteria} This subsection details how the initial candidate cohort of 195 participants was curated into the final benchmark of 183 participants and 5,224 validated 4~s PPG windows. The data curation protocol (Fig.~\ref{fig10}) was pre-specified a priori, adapted from \citet{c:50}, and aimed to: (i) ensure referential validity of reference capillary blood glucose and blood pressure labels, (ii) enforce hardware acquisition conformity on raw 40~s PPG recordings, (iii) remove only irrecoverable waveform corruption, and (iv) prevent bias by retaining clinically realistic yet difficult signals. We deliberately preserved moderate motion, respiration-induced baseline modulation, and perfusion variability to avoid an over-sanitized benchmark.
	
	\paragraph{Stage 1: Data Integrity Screening.}
	A subject was excluded only if one or more blood glucose or blood pressure labels were structurally missing or invalid. Any value outside the device’s nominal measurable range, such as a glucose reading beyond 0.6--33.3~mmol/L, was deemed invalid due to likely transcription or sensor failure and triggered exclusion after log confirmation. Values within the nominal range but outside 2--25~mmol/L were classified as suspicious anomalies and each was cross-checked against acquisition logs. If confirmed as a real physiological reading such as acute hyperglycemia, it was retained. If determined to be a recording or entry error, the subject was removed. Analogous plausibility validation was applied to systolic/diastolic pressures (systolic outside 50--250~mmHg, diastolic outside 30--150~mmHg treated as invalid unless log evidence supported authenticity). Three subjects were removed at this stage.
	
	\paragraph{Stage 2: Data Availability Screening.}
	PPG was acquired at a target 1000~Hz sampling rate in contiguous 40~s blocks (40,000 samples). We computed the realized sampling rate as $\hat{S} = (N-1)/\Delta t$ and enforced $\frac{|\hat{S}-1000|}{1000} \leq 0.005$. Block duration had to satisfy $|T-40.0|\leq 0.4$~s, a tolerance of $1\%$. Deviations beyond these hardware tolerance bounds indicated firmware or buffering errors rather than physiological variability and led to subject-level removal when systematic, affecting 2 subjects in total.
	
	\paragraph{Stage 3: Waveform Signal Quality Evaluation.}
	Following prior studies \citep{c:59,c:60,c:50}, we adopted a skewness-based index. Let $x=\{x_i\}_{i=1}^{N}$ denote the amplitude series for subject $s$, with $N=40{,}000$ samples for a single segment at 1000~Hz. Here $\hat{\mu}_x$ and $\sigma_x$ denote the sample mean and standard deviation. The segment-level skewness score is given by Eq.~\myref{eq:sqi_seg}:
	\begin{equation}
		{S_{\text{SQI}}}_{\text{seg}} = \frac{1}{N}\sum_{i=1}^{N}\left(\frac{x_i - \hat{\mu}_x}{\sigma_x}\right)^3.
		\label{eq:sqi_seg}
	\end{equation}
	If multiple valid 40~s segments existed for subject $s$, we computed each ${S_{\text{SQI}}}_{\text{seg},k}$ and defined the subject-level signal quality index as specified in Eq.~\myref{eq:sqi_subject}:
	\begin{equation}
		{S_{\text{SQI}}}= \frac{1}{K_s}\sum_{k=1}^{K_s}{S_{\text{SQI}}}_{\text{seg},k}.
		\label{eq:sqi_subject}
	\end{equation}
	Canonical pulsatile morphology (steep systolic ascent, gradual diastolic decay) yields a characteristic skew. Pervasive low pulsatility, sensor loosening, or amplitude collapse reduces the skew toward zero. We applied the pre-registered exclusion rule: remove subjects with ${S_\text{SQI}} < \mu_S$, where $\mu_S$ is the mean ${S_\text{SQI}}$ across all candidate subjects after Stages 1--2. This single, interpretable criterion avoids multi-threshold tuning and limits removals to clearly degraded acquisitions. Seven subjects fell below $\mu_S$ and were excluded.
	\paragraph{Stage 4: Time-window Segmentation.}
	After subject-level filtering, each retained raw 40~s segment was partitioned into non-overlapping 4~s windows (4,000 samples each), yielding 5{,}230 windows across the remaining 183 subjects.
	
	\paragraph{Stage 5: Short-Duration Flatline Screening.}
	To eliminate only windows devoid of usable pulsatile content, we marked as flatline any window containing a contiguous interval $\geq 0.5$~s with variance below $10^{-5}$ after z-score normalization of the window. In total, 6 windows accounting for $0.11\%$ of all windows were removed, primarily reflecting brief sensor contact loss. Parent subjects remained, yielding 5,224 final windows.
	\paragraph{Bias and Representativeness Assessment.}
	During subject-level curation, only $12$ of the initial $195$ participants ($6.15\%$) were excluded. To evaluate whether these removals could bias the dataset, we compared excluded and retained cohorts using tests matched to variable type. For continuous variables, reference glucose distributions showed no detectable shift in shape or location (two-sample Kolmogorov–Smirnov: $D=0.046$, $p=0.72$), and both age (Welch’s $t=0.58$, $p=0.56$, Cohen’s $d=0.08$) and systolic blood pressure (Welch’s $t=0.44$, $p=0.66$, $d=0.06$) differences were statistically non-significant with negligible effect sizes. For categorical variables, sex (Fisher’s exact: $p=0.68$, Cramér’s $V=0.05$) and diabetes-status proportions ($p=0.74$, $V=0.04$) were comparable between groups. These results, together with the minimal exclusion rate, support the conclusion that removals addressed unrecoverable acquisition faults rather than systematically eliminating atypical but valid physiological profiles, thereby preserving dataset representativeness.
	
	\paragraph{Rationale Against Over-Cleansing.}
	We intentionally avoided aggressive denoising or dynamic time-warping rejection rules that would preferentially enrich ``easy'' morphologies. Stages 1--5 constitute a conservative filter: retaining physiological and environmental variability (motion artifacts, temperature-related perfusion changes) while excising data that either (a) cannot be algorithmically rehabilitated or (b) plausibly represent instrumentation failure. This balances ecological validity with reproducibility and prevents artificial inflation of performance metrics.

	\paragraph{Anonymization.}
	Before archival, all direct identifiers (names, contact information, visit timestamps granular to the minute) were removed or coarse-grained. Encounter dates were shifted by a random, subject-specific offset held constant per subject, to preserve temporal spacing while protecting privacy.
	
	This principled pipeline yields the final curated benchmark of 183 subjects and 5{,}224 validated four-second windows, forming the evaluation substrate for the dynamic incremental learning framework without systematically excluding clinically challenging yet informative signal content.

	\subsection{PPG Preprocessing} \label{sec:ppg_preprocessing} To avoid ambiguity between dataset curation and model-input preprocessing, we separate the two steps explicitly. Section~3.1.2 first screens raw 40~s PPG recordings at the subject and recording levels. Only after these curation checks are the retained 40~s recordings partitioned once into ten non-overlapping 4~s windows, as described in Stage~4 of Fig.~\ref{fig10}. The present subsection describes the operations applied to these retained 4~s windows before model input. Thus, preprocessing does not introduce an additional segmentation stage beyond the one used to construct the benchmark windows. The same preprocessing pipeline was applied after the subject-level fold partition and only to recordings belonging to the corresponding fold. This fold-specific design prevents cross-subject statistics from being shared across training and held-out test subjects. In both training and test folds, each retained 4~s window was processed independently by physiologically motivated band-pass filtering and per-window normalization \citep{c:6,c:11,c:21,c:53}, as summarized in Fig.~\ref{fig:leakage_free_protocol}. \paragraph{Band-pass Filtering.} A zero-phase, forward--backward 4th-order Butterworth filter with passband 0.5--8~Hz was applied to each retained 4~s window. This range preserves the heart-rate fundamental and primary harmonics while attenuating sub-0.5~Hz baseline drift and sensor or quantization noise above 8~Hz. \paragraph{Normalization.} Filtered samples $x(t)$ inside each window were standardized as $\tilde{x}(t)=\big(x(t)-\mu_w\big)/\sigma_w, $ where $\mu_w$ and $\sigma_w$ denote the mean and standard deviation of that window. This local z-scoring stabilizes amplitude dynamics without using cross-window, cross-encounter, or cross-subject statistics. No additional baseline correction, motion-artifact suppression, wavelet transform, empirical mode decomposition, adaptive detrending, or template-based cleaning was applied, thereby maintaining reproducibility and avoiding step-specific hyperparameters.
	\subsection{Problem Definition}
	The proposed paradigm for physiological parameter measurement aims to learn from an evolving data stream with new distributions. The objective is to continuously build a regression model capable of handling the complete target space $\mathcal{Y}$. This data stream originates from both meta-scenarios $\mathcal{O}$ and real-world scenarios $\mathcal{R}$. We consider a set of \(B\) incremental training tasks \(\mathcal{T} = \left\{\mathcal{D}^1, \mathcal{D}^2, \cdots, \mathcal{D}^B\right\}\). The training instances for the \(b\)-th task are \(\mathcal{D}^b = \left\{\left(\mathbf{x}_i^b, y_i^b\right)\right\}_{i=1}^{n_b}\), where \(n_b\) is the number of instances. Each task's data are a union of data from the two scenarios, denoted as \(\mathcal{D}^b = \mathcal{D}_\mathcal{O}^b \cup \mathcal{D}_\mathcal{R}^b\), with \(\mathcal{D}_\mathcal{O}^b \neq \varnothing \vee \mathcal{D}_\mathcal{R}^b \neq \varnothing\).
	
	Specifically, the input \(\mathbf{x}_i^b \in \mathbb{R}^D\) is a one-dimensional PPG signal, where the dimension \(D\) is the product of the sampling duration \(T\) and the sampling rate \(S\). The corresponding output \(y_i^b \in \mathbb{R}\) is the blood glucose level, which serves as the regression target. We assume that training instances within each incremental step \(b\) are locally independent and identically distributed (IID), meaning each sample is drawn from the same underlying distribution without dependence on other samples in the step. This probabilistic relationship is formalized in Eq.~\myref{eq:1}:
	\begin{equation}
		P\left(\mathbf{x}_i^b,y_i^b|\mathcal{D}^b\right)=P\left(\mathbf{x}_i^b|\mathcal{D}^b\right)\cdot P\left(y_i^b|\mathbf{x}_i^b,\mathcal{D}^b\right).
		\label{eq:1}
	\end{equation}
	The model only accesses the training instances \(\mathcal{D}^b\) during task \(b\). Since \(y_i^b\in\mathbb{R}\) is a continuous blood glucose value, we do not define semantic classes over the target space. For clarity, we use \(\mathcal{Y}^{\leq b}=\{y_i^k \mid k=1,\ldots,b,\ i=1,\ldots,n_k\}\subset\mathcal{Y}\) to denote the continuous target values observed up to task \(b\). The task index \(b\) therefore refers to a ME2AC-discovered latent physiological task, rather than to a blood glucose class or a discretized target label.
	
	\paragraph{Protocol and fold isolation.} In all cross-validation runs, task discovery, episodic-memory construction, and proxy-gradient learning for unlabeled PPG signals (described in \cref{Dynamic Incremental Learning,MERDC,PGBA}) are performed using training-fold data only. The held-out test fold is excluded from clustering, memory exemplar selection, and any parameter update.
	\subsection{Dynamic Incremental Learning}
	\label{Dynamic Incremental Learning}
	The proposed Dynamic Incremental Learning (DIL), structure decomposes into a shared embedding module \(\phi^b(\cdot): \mathbb{R}^D \rightarrow \mathbb{R}^d\) and two independent linear layers. These layers are defined by learnable parameters \(\mathbf{w}_y^b\) and \(\mathbf{w}_\sigma^b \in \mathbb{R}^d\). The module \(\phi^b(\cdot)\) extracts a high-dimensional feature vector \(\mathbf{h} \in \mathbb{R}^d\) from an input instance \(\mathbf{x} \in \mathbb{R}^D\) to reflect its semantic information. This design is compatible with any backbone network architecture, such as Multilayer Perceptrons, Convolutional Neural Networks, or Vision Transformers \citep{c:20}. Based on previous findings \citep{c:15}, we use InceptionTime \citep{c:51} as the DIL backbone, a network known for its superior performance and resource efficiency in PPG-based physiological measurement tasks.
	
	The model produces a joint prediction \(\mathbf{J}(\mathbf{x})\) that includes both the target value prediction \(\hat{y}\) and its corresponding uncertainty estimate \(\hat{\sigma}^2\), as described in Eq.~\myref{eq:2}:
	\begin{equation}  
		\mathbf{J}(\mathbf{x}) = \begin{pmatrix}  
			\hat{y} \\
			\hat{\sigma}^2  
		\end{pmatrix} = \begin{pmatrix}  
			{\mathbf{w}_y^b}^\top \cdot \phi^b(\mathbf{x}) \\
			\exp({\mathbf{w}_\sigma^b}^\top \cdot \phi^b(\mathbf{x}))
		\end{pmatrix}.
		\label{eq:2}
	\end{equation}  
	Therefore, DIL aims to fit a model \(f(\mathbf{x}): X \rightarrow \mathbb{R} \times \mathbb{R}\) that returns both a blood glucose estimate and an auxiliary variance estimate. The regression branch is optimized by minimizing the expected squared prediction error over the incremental training distributions, as presented in Eq.~\myref{eq:3}:
	\begin{equation}
		\theta_y^*
		=
		\arg\min_{\theta\in\mathcal{H}}
		\mathbb{E}_{(\mathbf{x},y)\sim\mathcal{D}_t^1\cup\cdots\cup\mathcal{D}_t^b}
		\left[
		\left(y-\hat{y}_{\theta}(\mathbf{x})\right)^2
		\right],
		\label{eq:3}
	\end{equation}
	where \(\mathcal{H}\) is the hypothesis space and \(\mathcal{D}_t^b\) is the data distribution of task \(b\). The variance branch in Eq.~\myref{eq:2} is retained as a reliability-related output and is evaluated through the likelihood-based calibration procedure described below. This separation keeps the incremental regression objective consistent with the reported point-estimation metrics while allowing the uncertainty output to be assessed probabilistically.
	
	To mitigate catastrophic forgetting, we follow the gradient-episodic-memory paradigm from \citep{c:22}, but adapt its memory allocation rule to continuous blood glucose regression. We define a bounded episodic memory set \(\mathcal{E}=\{(\mathbf{x}_j,y_j)\}_{j=1}^{C_{\mathrm{mem}}}\) with a fixed capacity \(C_{\mathrm{mem}}\), where \(y_j\in\mathbb{R}\) remains a continuous regression label. Importantly, the memory budget is not allocated according to glucose classes or unique target values. Instead, at a given incremental update, ME2AC provides a task set \(\mathcal{T}=\{\mathcal{D}^1,\ldots,\mathcal{D}^B\}\), where \(B=|\mathcal{T}|\) is the number of discovered latent physiological tasks in the corresponding training fold. We allocate a task-balanced replay budget \[ m_b = \min\left(n_b,\left\lfloor\frac{C_{\mathrm{mem}}}{B}\right\rfloor\right), \qquad b=1,\ldots,B, \] and construct \(\mathcal{M}_b\subseteq\mathcal{D}^b\) with \(|\mathcal{M}_b|=m_b\). Within each discovered task, exemplars are selected using the same fold-local representation used by ME2AC so that the retained samples preserve feature-space diversity while keeping their original continuous blood glucose labels. Thus, replay is organized by unsupervised task partitions and feature-space representativeness, rather than by discretizing the regression target into classes. Consider a model parameterized by \(\theta \in \mathbb{R}^p\). The replay loss for task \(b\) is defined consistently with the regression objective in Eq.~\myref{eq:3}:
	\begin{equation}
		\ell(\theta,\mathcal{M}_b)
		=
		\frac{1}{|\mathcal{M}_b|}
		\sum_{(\mathbf{x}_i,y_i)\in\mathcal{M}_b}
		\left(y_i-\hat{y}_{\theta}(\mathbf{x}_i)\right)^2 .
		\label{eq:4}
	\end{equation}

	Minimizing only the current-task loss risks overfitting to recently observed data and thereby forgetting earlier tasks. To avoid this, we optimize the current task under explicit non-increase constraints on the replay memories of past tasks. Let \(\ell(\theta,\mathcal{D}^b)\) denote the empirical loss on the current task, defined analogously to Eq.~\myref{eq:4}. The constrained problem is then written as Eq.~\myref{eq:5}:
	\begin{equation}
		\begin{aligned}
			&\mathrm{minimize}_\theta\quad \ell(\theta,\mathcal{D}^b),\\
			&\mathrm{subject~to}\quad \ell(\theta,\mathcal{M}_k)\leq\ell(\theta^{b-1},\mathcal{M}_k)\text{ for all }k<b.
		\end{aligned}
		\label{eq:5}
	\end{equation}
	
	Here, \(\theta^{b-1}\) denotes the parameter state after completing task \(b-1\).
	
	Directly checking Eq.~\myref{eq:5} after every tentative update is inconvenient. Following gradient episodic memory, we therefore use a first-order surrogate based on gradient geometry. Let \(g=\nabla_\theta \ell(\theta,\mathcal{D}^b)\) be the raw gradient on the current task and \(g_k=\nabla_\theta \ell(\theta,\mathcal{M}_k)\) the memory gradient for past task \(k\). A candidate update direction \(\widetilde{g}\) preserves the loss on \(\mathcal{M}_k\) to first order if:
	\begin{equation}  
		\scalebox{1.15}{$
			\langle \widetilde{g},g_k\rangle \geq 0,\text{for all }k<b.
			$} 
		\label{eq:6}
	\end{equation}  
	
	If the raw gradient \(g\) violates this condition for any past task \(k\), we project \(g\) onto the closest feasible gradient \(\widetilde{g}\) that satisfies all constraints. This projection yields the update direction used to mitigate forgetting, as shown in Eq.~\myref{eq:7}:
	\begin{equation}  
		\begin{aligned}&\mathrm{minimize}_{\tilde{g}}\quad\frac{1}{2}\|g-\tilde{g}\|_2^2\\&\mathrm{subject~to}\quad\langle\tilde{g},g_k\rangle\geq0\text{ for all }k<b.\end{aligned}
		\label{eq:7}
	\end{equation}
	\paragraph{Uncertainty evaluation}
	The auxiliary variance branch provides an input-dependent scale for assessing the reliability of the blood glucose estimate. Because the main training objective is centered on point estimation, we evaluate this scale using a lightweight likelihood calibration within each subject-independent cross-validation cycle. Specifically, after the DIL model for the current outer cycle is trained, the predicted mean \(\hat{y}\) is kept fixed and the raw variance output is mapped to a calibrated log-variance:
	\begin{equation}
		s_i^{\mathrm{cal}}
		=
		a_k\log\left(\hat{\sigma}^{2}_{\mathrm{raw},i}+\epsilon\right)+b_k,
		\qquad
		\hat{\sigma}^{2}_{\mathrm{cal},i}
		=
		\exp\left(s_i^{\mathrm{cal}}\right),
		\label{eq:variance_calibration}
	\end{equation}
	where \(a_k\) and \(b_k\) are fold-specific calibration parameters for the \(k\)-th outer cycle, and \(\epsilon=10^{-8}\) is used for numerical stability. These two parameters are estimated from the four training folds by minimizing the Gaussian negative log-likelihood:
	\begin{equation}
		\mathcal{L}_{\mathrm{NLL}}
		=
		\frac{1}{2|\mathcal{D}^{k}_{\mathrm{train}}|}
		\sum_{i\in\mathcal{D}^{k}_{\mathrm{train}}}
		\left[
		\log(2\pi)
		+
		s_i^{\mathrm{cal}}
		+
		\frac{\left(y_i-\hat{y}_i\right)^2}{\exp\left(s_i^{\mathrm{cal}}\right)}
		\right].
		\label{eq:nll_calibration}
	\end{equation}
	The calibrated variance is used only for uncertainty-related reporting, including Gaussian NLL and prediction-interval coverage. The blood glucose estimates \(\hat{y}\) are not changed by this calibration step; therefore, the MAE, RMSE, Clarke Error Grid, and ISO-related point-estimation analyses are computed from the same frozen predictions as the main evaluation.
	\subsection{Mutual Entropy-Optimized Replay-Based Dynamic Clustering}
	\label{MERDC}
	\subsubsection{Method Overview}
	The MERDC module manages task discovery and replay under dynamic data streams, as outlined in Algorithm \ref{alg:1}. Its clustering core is the Mutual Entropy-optimized Enhanced Adaptive Clustering (ME2AC) operator. The process begins by applying ME2AC to the meta-scenario dataset \(\mathcal{D}_\mathcal{O}\). This initial step generates the task set \(\mathcal{T}\) and determines the number of clusters \(B\). Following this, the regression model \(f(x)\) undergoes training through the dynamic incremental learning paradigm. During this phase, episodic memories \(\mathcal{M}_b\) are sampled and stored for each task \(b\), forming a comprehensive memory set \(\mathcal{E}\).
	
	The next stage involves integrating real-world data. The episodic memory set \(\mathcal{E}\) merges with the real-scenario data \(\mathcal{D}_\mathcal{R}\) to create a new combined dataset \(\mathcal{D}_\mathcal{S}\). ME2AC is then reapplied within MERDC to update the task set and the number of clusters. This iterative cycle of model training, memory construction, and real-world data integration enables continuous optimization of the model. 

	\begin{algorithm}[t]
		\caption{Mutual Entropy-Optimized Replay-Based Dynamic Clustering}
		\label{alg:1}
		\begin{algorithmic}[1]
			\REQUIRE Training-fold meta-scenario data $\mathcal{D}_{\mathcal{O}}$, training-fold real-scenario data $\mathcal{D}_{\mathcal{R}}$, episodic memory capacity $C_{\mathrm{mem}}$
			\ENSURE Updated model $f(\mathbf{x})$
			
			\STATE \textbf{Step 1: Meta-clustering}
			\STATE $(\mathcal{T}, B) \leftarrow \mathrm{ME2AC}(\mathcal{D}_{\mathcal{O}})$ \hfill \textit{using training-fold PPG windows only}
			
			\STATE \textbf{Step 2: Model training}
			\STATE $(f(\mathbf{x}),\mathcal{E}) \leftarrow \mathrm{DIL}(\mathcal{T}, B, C_{\mathrm{mem}}, \widetilde{\mathcal{D}})$ \hfill \textit{see Algorithm~\ref{alg:2}}
			
			\STATE \textbf{Step 3: Episodic memory consolidation} \STATE $B \leftarrow |\mathcal{T}|$ \FOR{each discovered task $b \in \{1,\ldots,B\}$} \STATE $m_b \leftarrow \min\left(n_b,\left\lfloor C_{\mathrm{mem}}/B\right\rfloor\right)$ \STATE $\mathcal{M}_b \leftarrow \mathrm{Sample}_{\mathrm{div}}\left(\mathcal{D}^b,m_b\right)$ \hfill \textit{feature-diverse task-wise replay} \STATE $\mathcal{E} \leftarrow \mathcal{E} \cup \mathcal{M}_b$ \ENDFOR
			
			\STATE \textbf{Step 4: Real-world integration}
			\STATE $\mathcal{D}_{\mathcal{S}} \leftarrow \mathcal{E} \cup \mathcal{D}_{\mathcal{R}}$
			\STATE $(\mathcal{T}, B) \leftarrow \mathrm{ME2AC}(\mathcal{D}_{\mathcal{S}})$ \hfill \textit{using training-fold memory and real-scenario data only}
			
			\STATE \textbf{Step 5: Repeat}
			\STATE Repeat Steps 2--4 for subsequent incremental updates.
			
		\end{algorithmic}
	\end{algorithm}

	Algorithm~\ref{alg:1} is executed independently within each outer subject-independent 5CV cycle. In each cycle, the meta-scenario dataset $\mathcal{D}_\mathcal{O}$, the real-scenario dataset $\mathcal{D}_\mathcal{R}$, the combined dataset $\mathcal{D}_\mathcal{S}$, the task set $\mathcal{T}$, the number of discovered tasks $B$, the task-wise data $\mathcal{D}^b$, the episodic memory set $\mathcal{E}$, and the task memory subsets $\mathcal{M}_b$ are all constructed or updated using only subjects from the four non-test folds. Accordingly, all ME2AC-related quantities, including standardization statistics, histogram ranges, entropy estimates, mutual-information estimates, and clustering thresholds, are computed within the corresponding non-test folds and are not fitted using held-out test subjects. Likewise, the unlabeled PPG set $\widetilde{\mathcal{D}}$ used by PGBA, the proxy gradients $\bar{g}_i$, the projected gradients $\widetilde{g}$, and all parameter updates for $\theta$ and $\delta$ are computed only from non-test-fold data. The held-out test fold is not visible to any training-time operation, including ME2AC task-set generation, MERDC task-set updating, DIL sequential task-wise training, episodic memory construction, replay sampling, PGBA optimization, limited-label selection, model selection, hyperparameter tuning, or parameter updating; it is accessed only after the final trained model has been frozen for test inference and metric computation.
	\subsubsection{Mutual Entropy-Optimized Enhanced Adaptive Clustering}
	\label{Mutual Entropy-Optimized Enhanced Adaptive Clustering}
	
	We propose a novel unsupervised clustering operator named Mutual Entropy-optimized Enhanced Adaptive Clustering (ME2AC), which serves as the clustering core within MERDC. Given a set of preprocessed PPG windows \(\mathcal{D}=\{\mathbf{x}_1,\mathbf{x}_2,\ldots,\mathbf{x}_N\}\) from the training stream, ME2AC determines the number of discovered training tasks \(B\) and the corresponding task set \(\mathcal{T}\). The operator extends the DBSCAN framework by introducing two mechanisms: local signal entropy for core-point detection and histogram-based mutual information for cluster merging. The adaptive neighborhood radius further allows ME2AC to handle non-uniform local densities in longitudinal PPG streams.
	
	The feature space and the information-estimation procedure are clarified as follows. ME2AC uses preprocessed PPG windows for neighborhood construction, but it does not estimate mutual information as a continuous density in the original high-dimensional PPG waveform space. Instead, the entropy and mutual-information terms are computed as histogram-based plug-in estimates over discrete equivalence classes induced by binning the standardized PPG-window values. Samples or signal values falling into the same histogram bin are treated as belonging to the same empirical equivalence class. Therefore, the information-theoretic quantities used by ME2AC should be interpreted as discrete empirical estimates of local morphology and cluster-level similarity patterns, rather than direct high-dimensional continuous mutual-information estimates.
	
	The process begins with fold-local standardization of the preprocessed PPG windows. For the \(i\)-th window \(\mathbf{x}_i=(x_{i1},x_{i2},\ldots,x_{iD})\), the standardized point \(\mathbf{z}_i=(z_{i1},z_{i2},\ldots,z_{iD})\) is calculated as
	\begin{equation}
		z_{ij}
		=
		\frac{x_{ij}-\mu_j}{\sigma_j+\xi},
		\quad
		j=1,2,\ldots,D,
		\label{eq:8}
	\end{equation}
	where \(\mu_j\) and \(\sigma_j\) denote the mean and standard deviation of the \(j\)-th waveform coordinate computed from the current training-fold clustering set, and \(\xi=10^{-8}\) is used for numerical stability. This standardization is applied after the preprocessing steps in Section~\ref{sec:ppg_preprocessing}. No blood glucose labels, subject identifiers, demographic variables, validation subjects, or test subjects are used during ME2AC task discovery.
	
	Next, ME2AC calculates a dynamic neighborhood radius \(\epsilon_i\) for each standardized PPG point \(\mathbf{z}_i\). The radius is defined as the median distance to its \(k\)-nearest neighbors:
	\begin{equation}
		\epsilon_i
		=
		\operatorname{median}
		\left(
		\left\{
		\|\mathbf{z}_i-\mathbf{z}_j\|_2
		\mid
		j\in\operatorname{kNN}(\mathbf{z}_i)
		\right\}
		\right),
		\label{eq:9}
	\end{equation}
	where \(\operatorname{kNN}(\mathbf{z}_i)\) denotes the set of \(k\)-nearest neighbors of \(\mathbf{z}_i\). This step is used to define the local neighborhood structure and avoids the fixed-radius assumption of conventional DBSCAN. It does not by itself constitute a high-dimensional information-density estimate.
	
	The algorithm then performs core-point detection. For each standardized point \(\mathbf{z}_i\), ME2AC computes the local signal entropy \(H_i\) within its adaptive neighborhood \(\mathcal{N}_i=\{\mathbf{z}_j:\|\mathbf{z}_j-\mathbf{z}_i\|_2\le\epsilon_i\}\). The entropy calculation is based on a histogram-induced empirical distribution. Specifically, the standardized signal values from the neighborhood are assigned to \(K_H\) bins shared within the current training-fold clustering set. Let \(n_{im}\) denote the count of values in \(\mathcal{N}_i\) falling into the \(m\)-th bin. The local entropy is computed as
	\begin{equation}
		H_i
		=
		-\sum_{m=1}^{K_H}p_{im}\log(p_{im}+\xi),
		\quad
		p_{im}
		=
		\frac{n_{im}}{\sum_{r=1}^{K_H}n_{ir}},
		\label{eq:10}
	\end{equation}
	where \(p_{im}\) is the empirical probability of the \(m\)-th equivalence bin. A point \(\mathbf{z}_i\) is marked as a core point if \(H_i\le\tau_H\). The threshold \(\tau_H\) is fixed within each training fold before test inference and is examined in the sensitivity analysis reported in Section~\ref{sec:me2ac_sensitivity}.
	
	Clusters are then recursively expanded from the identified core points. For a core point \(\mathbf{z}_c\), all unclassified points within its adaptive neighborhood \(\mathcal{N}_c\) are assigned to the current cluster \(c\), denoted as \(\mathbf{L}(\mathcal{N}_c)\gets c\), where \(\mathbf{L}\) is the vector of cluster labels. This step follows the density-expansion principle of DBSCAN, while the core-point decision is governed by the local entropy criterion above.
	
	Finally, ME2AC merges highly related preliminary clusters using histogram-based mutual information. For a candidate cluster pair \((\mathcal{D}^k,\mathcal{D}^l)\), the same binning rule is used to construct empirical marginal and joint frequency tables over the histogram-induced equivalence classes. The mutual-information estimate is computed as
	\begin{equation}
		\operatorname{MI}(\mathcal{D}^k,\mathcal{D}^l)
		=
		\sum_{b_x=1}^{B}
		\sum_{b_y=1}^{B}
		\frac{n(b_x,b_y)}
		{|\mathcal{D}^k|+|\mathcal{D}^l|}
		\log
		\left(
		\frac{
			n(b_x,b_y)(|\mathcal{D}^k|+|\mathcal{D}^l|)
		}
		{
			|\mathcal{D}^k||\mathcal{D}^l|n(b_x)n(b_y)+\xi
		}
		\right),
		\label{eq:11}
	\end{equation}
	where \(n(b_x,b_y)\) is the joint histogram frequency of the \((b_x,b_y)\)-th equivalence-bin pair, and \(n(b_x)\) and \(n(b_y)\) are the corresponding marginal histogram frequencies for clusters \(\mathcal{D}^k\) and \(\mathcal{D}^l\), respectively. This formulation is a discrete histogram-based plug-in estimator and should not be interpreted as estimating a continuous joint density over the original PPG waveform space.
	
	If \(\operatorname{MI}(\mathcal{D}^k,\mathcal{D}^l)\ge\tau_{\mathrm{MI}}\), the two clusters are regarded as sufficiently related and are merged by setting \(\mathbf{L}(\mathcal{D}^l):=k\). The threshold \(\tau_{\mathrm{MI}}\) is fixed within each training fold before test inference and is examined in the sensitivity analysis reported in Section~\ref{sec:me2ac_sensitivity}. Overall, ME2AC performs unsupervised task discovery from standardized PPG-window distributions. It uses neither blood glucose labels nor subject-level metadata during clustering, and all standardization statistics, histogram ranges, entropy thresholds, and mutual-information thresholds are obtained without using the held-out test fold.
	\subsection{Proxy Gradient Bridging Agent} \label{PGBA} The objective of Dynamic Incremental Learning (DIL), as defined in Eqs.~\myref{eq:3}--\myref{eq:7}, is to minimize the regression risk under a non-stationary data stream while preserving previously learned task knowledge. For a labeled sample \((\mathbf{x}_i,y_i)\), the standard supervised update is obtained by the chain rule: \begin{equation} \theta\leftarrow\theta-\eta \frac{\partial\ell}{\partial\mathbf{J}(\mathbf{x}_i)} \frac{\partial\mathbf{J}(\mathbf{x}_i)}{\partial\theta}, \label{eq:12} \end{equation} where \(\eta\) is the learning rate and \(\mathbf{J}(\mathbf{x})=(\hat{y},\hat{\sigma}^{2})^\top\) denotes the joint output of the DIL model. This update cannot be directly applied to an unlabeled PPG window \(\widetilde{\mathbf{x}}_i\in\widetilde{\mathcal{D}}\), because the supervised output-space gradient \(\partial\ell/\partial\mathbf{J}\) depends on the unavailable blood glucose label. To leverage unlabeled PPG streams without assigning pseudo glucose labels, we introduce a Proxy Gradient Bridging Agent (PGBA), denoted by \(a(\cdot;\delta)\), where \(\delta\) represents the learnable agent parameters. PGBA maps the current model output to an output-space proxy gradient: \begin{equation} \bar{g}_i=a(\mathbf{J}(\mathbf{x}_i);\delta) \approx \frac{\partial\ell}{\partial\mathbf{J}(\mathbf{x}_i)} . \label{eq:pgba_mapping} \end{equation} For an unlabeled window \(\widetilde{\mathbf{x}}_i\), the predicted proxy gradient \(\bar{g}_{\widetilde{x}_i}=a(\mathbf{J}(\widetilde{\mathbf{x}}_i);\delta)\) substitutes for the unavailable supervised output gradient, yielding the proxy-gradient update: \begin{equation} \theta\leftarrow\theta-\eta\, \bar{g}_{\widetilde{x}_i} \frac{\partial\mathbf{J}(\widetilde{\mathbf{x}}_i)}{\partial\theta}. \label{eq:13} \end{equation} Thus, unlabeled PPG windows contribute through output-space update directions rather than through pseudo-label regression targets. For labeled samples, the true supervised gradient direction is available. PGBA is trained by a bridge loss that evaluates whether a proxy step in the output space would reduce the supervised loss: \begin{equation} \ell_{\mathrm{bridge}}(\mathbf{J}(\mathbf{x}_i),\bar{g}_i,y_i) = \ell(\mathbf{J}(\mathbf{x}_i)-\eta\bar{g}_i,y_i). \label{eq:14} \end{equation} The agent parameters are optimized by minimizing this bridge loss: \begin{equation} \delta\leftarrow\delta-\hat{\eta}\, \frac{\partial\ell_{\mathrm{bridge}}}{\partial\bar{g}_i} \frac{\partial\bar{g}_i}{\partial\delta}. \label{eq:15} \end{equation} The bridge loss provides a local explanation for why PGBA encourages useful proxy directions. Let $ q_i={\partial\ell}/{\partial\mathbf{J}(\mathbf{x}_i)} $ be the true supervised output-space gradient. If \(\ell\) is locally \(L\)-smooth with respect to \(\mathbf{J}\), then a one-step proxy update satisfies \begin{equation} \ell(\mathbf{J}_i-\eta\bar{g}_i,y_i) \leq \ell(\mathbf{J}_i,y_i) -\eta\langle q_i,\bar{g}_i\rangle + \frac{L\eta^2}{2}\|\bar{g}_i\|^2 . \label{eq:pgba_alignment} \end{equation} Therefore, minimizing the bridge loss is locally favored when \(\langle q_i,\bar{g}_i\rangle>0\), i.e., when the proxy gradient is aligned with the supervised output gradient. PGBA is not required to exactly reproduce the full supervised gradient magnitude; it is trained to provide a locally useful update direction for unlabeled windows. To prevent excessive proxy-gradient magnitudes, the predicted proxy gradient is normalized before use: \begin{equation} \bar{g}_i = \alpha\varphi_i \frac{a(\mathbf{J}(\mathbf{x}_i);\delta)} {\|a(\mathbf{J}(\mathbf{x}_i);\delta)\|+\xi}, \qquad \varphi_i= \left\| \frac{\partial\ell}{\partial\mathbf{J}(\mathbf{x}_i)} \right\|, \label{eq:16} \end{equation} where \(\alpha\in(0,1)\) is a scaling factor and \(\xi=10^{-8}\) avoids numerical instability. This gives \(\|\bar{g}_i\|\leq\alpha\varphi_i\), thereby bounding the proxy step by a fixed fraction of the supervised output-gradient magnitude observed on labeled samples. This bounded-update property is the main safeguard used in PGBA to reduce the risk of biased or overly large proxy updates. The complete DIL paradigm integrating PGBA is summarized in Algorithm~\ref{alg:2}. In the experiments, PGBA is used only within the training folds. Held-out test subjects are never used to train the agent, compute proxy gradients, or tune its parameters.
	
	As illustrated in Fig. \ref{fig2}(b), knowledge derived from non-stationary data distributions progressively enhances the capacity of the latent representation space for blood glucose estimation. This, in turn, mitigates the uncertainty that arises from inter-subject variability. This workflow improves the adaptability of the estimation method across diverse scenarios and ensures its robustness and generalization in practical applications.
	\begin{algorithm}[t] \caption{Dynamic Incremental Learning with PGBA} \label{alg:2} \begin{algorithmic}[1] \REQUIRE ME2AC-discovered task set $\mathcal{T}=\{\mathcal{D}^1,\ldots,\mathcal{D}^B\}$, training-fold unlabeled set $\widetilde{\mathcal{D}}$, memory capacity $C_{\mathrm{mem}}$, model parameters $\theta$, PGBA parameters $\delta$ \ENSURE Updated model $f_\theta$ and episodic memory $\mathcal{E}$ \STATE Initialize $\mathcal{E}\leftarrow\varnothing$ and $\mathcal{M}_b\leftarrow\varnothing$ for all \(b=1,\ldots,B\) \STATE Allocate task-wise memory budget \(m_b\leftarrow\min(n_b,\lfloor C_{\mathrm{mem}}/B\rfloor)\) \FOR{$b=1,\ldots,B$} \FOR{each labeled mini-batch from $\mathcal{D}^b$} \STATE Compute supervised gradient \(g_{\mathrm{sup}}\) using Eq.~\myref{eq:12} \STATE Update task memory \(\mathcal{M}_b\) using \(\mathrm{UpdateDiv}(\cdot)\) \STATE Train PGBA on labeled samples using Eqs.~\myref{eq:14}--\myref{eq:16} \IF{$\widetilde{\mathcal{D}}\neq\varnothing$} \STATE Sample an unlabeled mini-batch from \(\widetilde{\mathcal{D}}\) \STATE Compute proxy-gradient update \(g_{\mathrm{pgba}}\) using Eq.~\myref{eq:13} \STATE Combine current updates \(g\leftarrow g_{\mathrm{sup}}+g_{\mathrm{pgba}}\) \ELSE \STATE \(g\leftarrow g_{\mathrm{sup}}\) \ENDIF \IF{$b>1$} \STATE Compute memory gradients \(g_k=\nabla_\theta\ell(\theta,\mathcal{M}_k)\), \(k<b\) \STATE Project \(g\) to \(\widetilde{g}\) under the non-interference constraints in Eq.~\myref{eq:7} \ELSE \STATE \(\widetilde{g}\leftarrow g\) \ENDIF \STATE Update model parameters \(\theta\leftarrow\theta-\eta\widetilde{g}\) \ENDFOR \STATE \(\mathcal{E}\leftarrow\mathcal{E}\cup\mathcal{M}_b\) \ENDFOR \RETURN \(f_\theta,\mathcal{E}\) \end{algorithmic} \end{algorithm}
	
	In Algorithm~\ref{alg:2}, the replay budget is allocated across ME2AC-discovered latent tasks rather than across blood glucose categories. Specifically, \(B=|\mathcal{T}|\) denotes the number of task partitions discovered within the current training fold, and \(m_b=\min(n_b,\lfloor C_{\mathrm{mem}}/B\rfloor)\) is the memory budget assigned to task \(b\). The operator \(\mathrm{UpdateDiv}\) maintains a task-wise memory subset by retaining representative samples in the fold-local ME2AC feature space. Blood glucose values are not discretized into semantic classes during this process; instead, they are preserved as continuous regression labels. Thus, the replay buffer is organized according to unsupervised task partitions and feature-space diversity. When multiple candidates provide comparable feature-space coverage, samples that improve the coverage of the continuous glucose range within the task are preferred.
\vspace{-1em}
	\section{Experiments}
	\label{sec:experiments}
	\subsection{Implementation Details}
	
	We conducted subject-independent five-fold cross-validation (5CV) over 183 subjects to ensure a robust and unbiased assessment of model performance. The cohort was partitioned into five mutually exclusive subject folds of sizes 37, 37, 37, 36, and 36, reflecting that 183 is not perfectly divisible by five. We performed five cross-validation cycles. In each cycle, one subject fold served as the held-out test fold and the remaining four folds formed the training folds. Thus, when a 37-subject fold was held out, the training folds contained 146 subjects, whereas when a 36-subject fold was held out, the training folds contained 147 subjects. This 5CV protocol averages performance across multiple train--test partitions, reduces variance due to any particular subject split, and evaluates generalization to unseen subjects. We considered leave-one-out cross-validation (LOOCV) for finer granularity, but rejected it because 183 training iterations would be computationally inefficient relative to the marginal precision gain for this dataset scale of 5,224 validated four-second windows.
	
	\begin{figure}[pos=htbp]
		\centering
		\includegraphics[width=\textwidth]{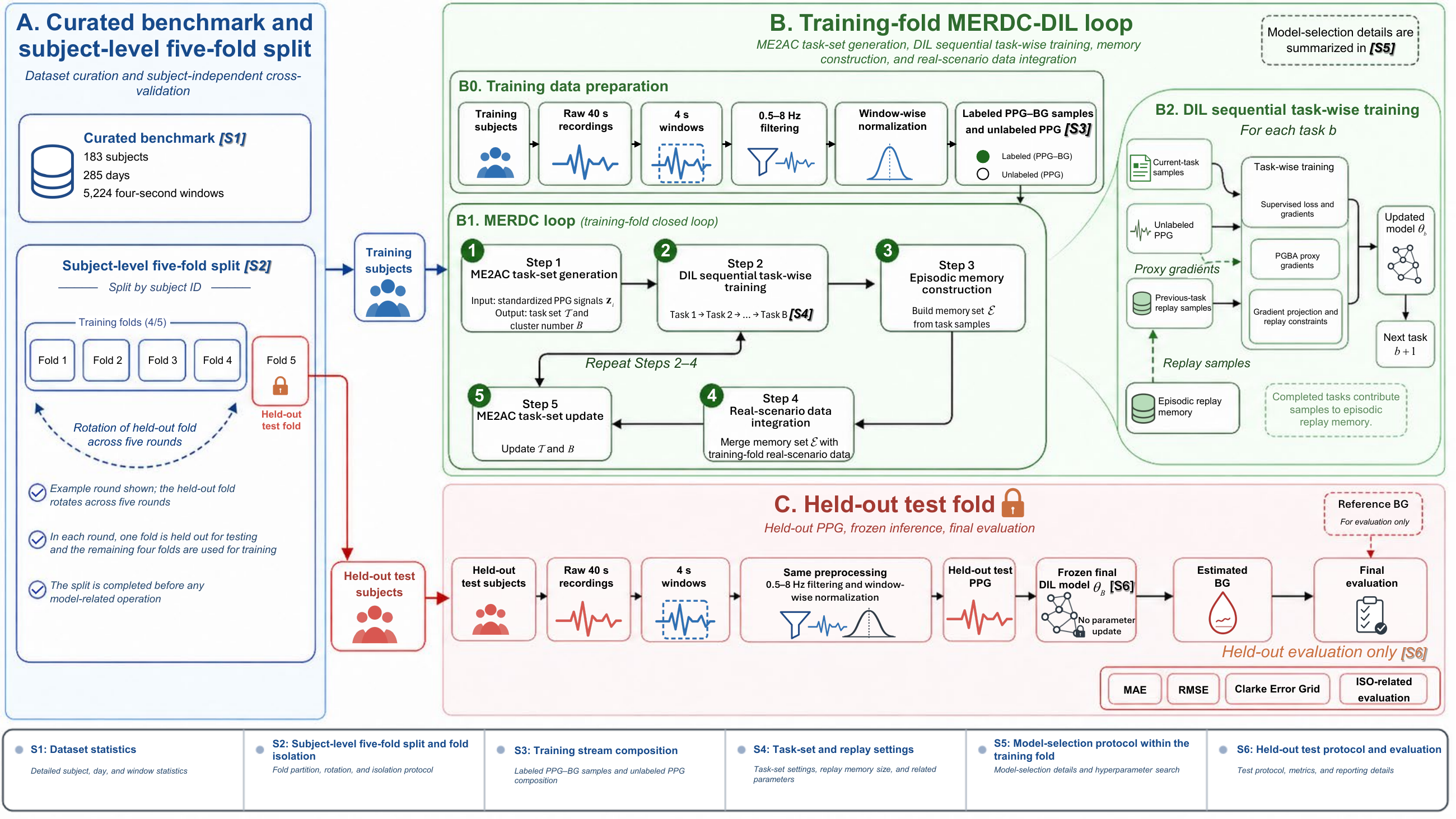}
		\caption{Leakage-free subject-independent five-fold cross-validation and training-fold MERDC-DIL pipeline. Participants were first partitioned into five mutually exclusive subject folds. In each cycle, one fold was held out for testing and the remaining four folds were used for training. All training-time operations, including ME2AC task-set generation, DIL sequential task-wise training, PGBA-assisted learning, historical replay, episodic memory construction, real-scenario data integration, and ME2AC task-set updating, were restricted to the training folds. The held-out test fold was used only for frozen inference and final evaluation. Reference BG values from held-out subjects were used only for evaluation and were never used for task discovery, memory construction, PGBA optimization, model selection, or parameter updating. S1--S6 denote the corresponding reporting items summarized in Table~\ref{tab:fold_isolation_protocol}.}
		\label{fig:leakage_free_protocol}
	\end{figure}
	
	Across the full benchmark, there are 4,104 four-second signal segments from non-diabetic subjects and 1,120 from subjects with diabetes. Within each cross-validation cycle, the subject-level partition was completed before any model-related operation. Subjects in the held-out test fold were used exclusively for frozen inference and final evaluation. They were never used in any training-time operation, including ME2AC task-set generation, MERDC task-set updating, DIL sequential task-wise training, episodic memory construction, replay-based optimization, PGBA training, limited-label selection, model-selection decisions, hyperparameter tuning, or parameter updating. The training-fold subjects were further organized into labeled and unlabeled streams: labeled PPG--BG samples provided supervised task-wise learning, whereas for the unlabeled stream only PPG windows were used and the corresponding BG labels were withheld. These unlabeled PPG windows contributed to training only through PGBA's uncertainty-quantified proxy gradients. Analyses of non-stationary real-world streams, including those illustrated in Fig.~\ref{fig4}, followed the same fold-isolated protocol: newly arriving data used for adaptation were restricted to the training folds, whereas held-out test subjects were reserved for frozen inference and final metric computation.
	
	To make the fold-wise data flow and leakage-control protocol explicit, Fig.~\ref{fig:leakage_free_protocol} summarizes the complete evaluation pipeline. In the training folds, raw 40~s recordings were segmented into 4~s windows and processed using the same preprocessing steps as in Section~3.2. ME2AC was applied only to training-fold standardized PPG signals to generate the task set $\mathcal{T}$ and cluster number $B$. DIL then performed sequential task-wise training over $\mathcal{T}$; during each task-wise update, current-task samples, training-fold unlabeled PPG through PGBA, and historical replay samples were used. Episodic memory was constructed only from training-subject task samples. The memory set $\mathcal{E}$ and training-fold real-scenario data were then integrated within MERDC to update $\mathcal{T}$ and $B$. In contrast, the held-out test fold followed only the preprocessing and frozen-inference path. Reference BG values from the held-out subjects were used only for final metric computation.

	Table~\ref{tab:fold_isolation_protocol} further specifies the data visibility, model-update status, and purpose of each stage. This table is intended to clarify that the proposed dynamic task discovery and replay mechanisms operate only within the training folds in every 5CV cycle.
	
	\begin{table}[pos=t]
		\centering
		\caption{Data visibility and fold isolation in the leakage-free subject-independent 5CV protocol.}
		\label{tab:fold_isolation_protocol}
		\footnotesize
		\setlength{\tabcolsep}{3.0pt}
		\renewcommand{\arraystretch}{1.15}
		\begin{tabularx}{\textwidth}{
				>{\raggedright\arraybackslash}p{0.17\textwidth}
				>{\raggedright\arraybackslash}p{0.25\textwidth}
				>{\raggedright\arraybackslash}p{0.17\textwidth}
				>{\raggedright\arraybackslash}p{0.14\textwidth}
				>{\raggedright\arraybackslash}X}
			\toprule
			Stage & Data used & Held-out test fold visible? & Model update? & Purpose \\
			\midrule
			Subject-level split & Subject IDs only & No signal or label exposure & No & Construct five mutually exclusive subject folds \\
			Training data preparation & Training-fold 40~s PPG recordings & No & No & Generate 4~s windows and apply filtering and normalization \\
			ME2AC task-set generation & Training-fold standardized PPG signals & No & No direct parameter update & Generate task set $\mathcal{T}$ and cluster number $B$ \\
			DIL sequential task-wise training & Training-fold labeled PPG--BG samples & No & Yes, training folds only & Supervised task-wise regression learning \\
			PGBA-assisted learning & Training-fold unlabeled PPG only & No & Yes, training folds only & Generate proxy gradients for unlabeled PPG \\
			Historical replay & Previous-task replay samples from training folds & No & Yes, training folds only & Mitigate forgetting during task-wise updates \\
			Episodic memory construction & Training-subject task samples & No & No direct parameter update & Build memory set $\mathcal{E}$ for subsequent MERDC integration \\
			Real-scenario data integration & Training-fold real-scenario data and $\mathcal{E}$ & No & Training folds only & Update task set through MERDC \\
			Fixed evaluation settings & Predefined/default configurations and training-fold protocol & No test exposure & No test update & Keep all method settings fixed across folds without test-fold tuning \\
			Held-out test inference & Held-out test PPG only & Yes, inference only & No & Frozen-model BG estimation \\
			Final evaluation & Estimated BG and reference BG from held-out subjects & Yes, evaluation only & No & Report MAE, RMSE, Clarke Error Grid, and ISO-related metrics \\
			\bottomrule
		\end{tabularx}
	\end{table}

To further support reproducibility without introducing fold-specific tuning, we explicitly report the fixed training-stream organization used in all subject-independent 5CV cycles. Within the four training folds of each cycle, 75\% of the training windows were used as labeled PPG--BG samples and the remaining 25\% were treated as unlabeled PPG-only samples by withholding their BG labels, resulting in a 3:1 labeled-to-unlabeled ratio. This split was performed only within the training folds. The held-out test fold was never used for limited-label selection, PGBA training, task discovery, hyperparameter adjustment, or model updating. 

The supervised and unlabeled streams were combined according to the fixed update schedule summarized in Algorithm~\ref{alg:2}. For each current-task labeled mini-batch, the model first computed the supervised regression gradient. PGBA was then updated through the bridge loss on labeled samples and, when an unlabeled mini-batch was available, produced a bounded proxy-gradient update for unlabeled PPG windows. Replay samples from previous tasks were used only after the corresponding episodic memory buffer had been constructed. Therefore, unlabeled PPG windows contributed to training through proxy gradients rather than through pseudo glucose labels. 

The task number was not pre-specified. ME2AC determined the fold-specific task set $\mathcal{T}$ and cluster number $B$ using only training-fold representations. Across the five 5CV cycles, the fold-wise task numbers were $B=\{5,5,5,5,6\}$, corresponding to $5.2\pm0.4$ tasks under the default ME2AC setting. For ME2AC, Euclidean distance was used for neighborhood construction. The default information-estimation setting was $k=5$, $K_H=10$, $K_{MI}=10$, $\alpha_H=0.60$, and $\alpha_{MI}=0.75$, where $\alpha_H$ and $\alpha_{MI}$ denote the fold-local quantile rules used to determine the entropy and mutual-information thresholds, respectively. The sensitivity of these settings is reported in Section~\ref{sec:me2ac_sensitivity}. 

All proposed-method settings were fixed before the final 5CV evaluation and kept unchanged across folds: the initial learning rate was \(1\times10^{-3}\), the PGBA agent learning rate was \(\hat{\eta}=1\times10^{-3}\), the batch size was 128, the episodic memory capacity was 512, the number of epochs per task was 20, the PGBA scaling factor was \(\alpha=0.15\), and the agent architecture was \(a(\cdot;\delta)=(128,32)\) with proxy-gradient normalization. The numerical stabilizer in Eq.~\myref{eq:16} was fixed as \(\xi=10^{-8}\). These settings were used as a fixed configuration rather than being selected separately for each fold, and no held-out test subject was used for any hyperparameter adjustment.
	
	\begin{figure}[pos=htbp]
		\centering
		\includegraphics[width=0.3\linewidth]{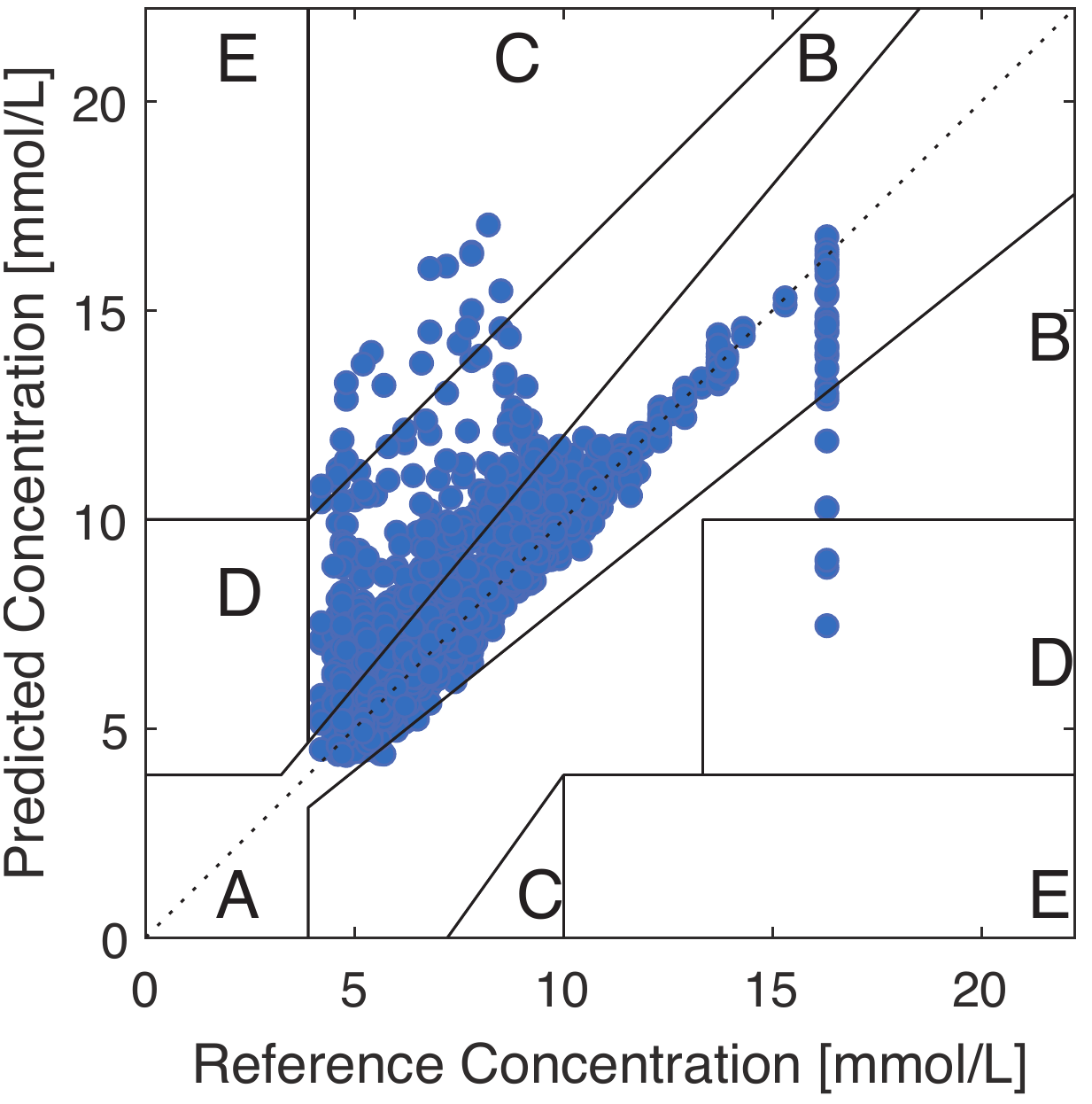}
		\caption{Clarke Error Grid analysis of the proposed MERDC-enhanced dynamic incremental learning framework under subject-independent evaluation. Each point represents a window-level blood glucose estimate paired with the corresponding reference capillary blood glucose measurement. The grid visualizes clinically relevant error regions, while quantitative comparisons with reproduced baseline methods are reported in Table~\ref{tab:2}, and aggregation-aware ISO-related analyses are reported in Table~\ref{tab:3} and Table~\ref{tab:slice_robustness}.}
		\label{fig13}
		\vspace{-2em}
	\end{figure}
	\subsection{Evaluation Criteria}
	To evaluate the performance of the proposed paradigm for blood glucose (BG) estimation, we follow mainstream metrics \citep{c:2,c:10,c:18,c:15,c:13,c:6}, including Mean Absolute Error (MAE) and Root Mean Square Error (RMSE) for accuracy assessment, the Clarke Error Grid (CEG) \citep{c:52} for categorizing the percentage of estimates across clinically relevant regions, and the ISO 15197:2013 standard \citep{c:27} for assessing the accuracy and reliability of BG measurement devices. 
	
	For ISO-related analysis, we used the ISO 15197:2013 reference error bands as measurement-quality criteria rather than as a regulatory-compliance claim. A prediction was counted as falling within the reference band if its absolute error was within $\pm 0.83$ mmol/L for reference BG below $5.55$ mmol/L, or within $\pm 15\%$ for reference BG at or above $5.55$ mmol/L. Because each 40-s PPG acquisition is divided into ten non-overlapping 4-s windows that share the same reference capillary BG measurement, we report these error-band statistics at multiple aggregation levels. Window-level analysis treats each 4-s window as one prediction unit. Acquisition-level analysis first averages the valid 4-s window predictions derived from the same 40-s acquisition and then compares the averaged prediction with the corresponding reference BG. Subject-macro analysis accounts for repeated acquisitions from the same participant by first computing subject-specific acquisition-level metrics and then averaging these metrics across subjects. Ninety-five percent confidence intervals were estimated by non-parametric bootstrap resampling over the corresponding aggregation unit.

	For uncertainty assessment, we additionally report Gaussian negative log-likelihood (NLL) and prediction interval coverage probability (PICP). The calibrated predictive interval is constructed from the point estimate \(\hat{y}_i\) and the calibrated standard deviation \(\hat{\sigma}_{\mathrm{cal},i}\) as \(\hat{y}_i \pm z_q\hat{\sigma}_{\mathrm{cal},i}\), where \(z_q\) denotes the standard normal quantile for a nominal two-sided coverage level \(q\). In the main results, we report 95\% PICP and the absolute coverage gap \(\mathrm{Gap}_{95}=|\mathrm{PICP}_{95}-0.95|\), which indicate whether the uncertainty interval is under- or over-confident on held-out subjects.
	\subsection{Comparisons with State-of-the-art Methods}
	Non-invasive blood glucose (BG) estimation based on photoplethysmography (PPG) faces significant challenges, including intra-individual data drift and external variability, which reduce the robustness and practical applicability of existing models. To validate the advantages of our paradigm, which integrates an uncertainty-quantified proxy gradient bridging agent and mutual entropy--optimized replay clustering within a dynamic incremental learning framework, we conducted a comprehensive comparison with state-of-the-art (SOTA) methods.
	
	\begin{table}[pos=htbp]  
	\centering
	\normalfont\footnotesize
	\setlength{\tabcolsep}{4pt}
	\caption{Comparison of different methods on the benchmark dataset.}
	\begin{threeparttable}
		\begin{tabularx}{\linewidth}{@{}>{\raggedright\arraybackslash}Xccc@{}}
			\toprule
			Method & MAE (mmol/L) $\downarrow$ & RMSE (mmol/L) $\downarrow$ & Zone A (\%) $\uparrow$ \\
			\midrule
			\cite{c:16} & 4.63 $\pm$ 0.49 & 6.67 $\pm$ 0.81 & 18.83 $\pm$ 1.30 \\
			\cite{c:10} & 8.35 $\pm$ 7.59 & 9.13 $\pm$ 7.41 & 21.92 $\pm$ 19.24 \\
			\cite{c:11} & 1.50 $\pm$ 0.17 & 2.58 $\pm$ 0.18 & 59.78 $\pm$ 6.58 \\
			\midrule
			\cite{c:36} & 1.42 $\pm$ 0.05 & 2.27 $\pm$ 0.09 & 60.71 $\pm$ 1.83 \\
			\cite{c:55} & 1.12 $\pm$ 0.03 & 1.76 $\pm$ 0.11 & 71.38 $\pm$ 1.47 \\
			\cite{c:12} & 1.45 $\pm$ 0.03 & 2.55 $\pm$ 0.12 & 64.89 $\pm$ 1.38 \\
			\cite{c:15} & \underline{0.99 $\pm$ 0.05} & 1.81 $\pm$ 0.14 & \underline{79.21 $\pm$ 1.90} \\
			\cite{c:37} & 1.19 $\pm$ 0.02 & 1.83 $\pm$ 0.09 & 68.70 $\pm$ 0.64 \\
			\cite{c:53} & 1.29 $\pm$ 0.02 & 2.48 $\pm$ 0.05 & 70.85 $\pm$ 1.17 \\
			\cite{c:54} & 1.06 $\pm$ 0.02 & \underline{1.67 $\pm$ 0.08} & 73.70 $\pm$ 1.91 \\
			\midrule
			\rowcolor{black!8}
			Ours & \textbf{0.64 $\pm$ 0.01} & \textbf{1.29 $\pm$ 0.10} & \textbf{86.48 $\pm$ 3.10} \\
			\bottomrule
		\end{tabularx}
	\end{threeparttable}
	\label{tab:2}
\end{table}
	
	We evaluated our approach on a newly constructed benchmark dataset comprising PPG signals and corresponding blood glucose (BG) values collected from 183 participants over 285 days. This dataset spans a mean participant age of $57.02 \pm 21.20$ years and BG levels of 4.2--16.3 mmol/L, ensuring demographic and glycaemic diversity. As shown in Table~\ref{tab:2}, under this unified fixed-configuration protocol, our method achieves a Mean Absolute Error (MAE) of $0.64 \pm 0.01$ mmol/L and a Root Mean Square Error (RMSE) of $1.29 \pm 0.10$ mmol/L, outperforming the reproduced baseline implementations.
	
	All baseline results in Table~\ref{tab:2} were obtained by re-implementing the respective methods on our benchmark under the same inclusion/exclusion rules and subject-independent 5CV protocol. To avoid unequal optimization or test-set leakage, we did not perform fold-specific or test-informed hyperparameter tuning for either the baselines or the proposed method. Baseline architectures, feature definitions, and training configurations followed the original publications or publicly available default settings whenever available. The proposed method was likewise evaluated using the fixed configuration reported in Section~4.1. Therefore, Table~\ref{tab:2} should be interpreted as a unified fixed-configuration benchmark comparison rather than as a post-hoc hyperparameter search over competing methods.
	
	Traditional methods \citep{c:16,c:10} rely on handcrafted features such as Gaussian-derived descriptors and shallow regressors such as Support Vector Regression (SVR), and incur high errors with MAE exceeding $4.6$ mmol/L while failing to address dynamic signal variation. Subsequent studies \citep{c:11,c:36} improved accuracy to approximately $1.5$ mmol/L in MAE by leveraging machine learning approaches such as Random Forest Regression (RFR) and multi-feature fusion, but lacked mechanisms to handle intra-individual data drift. Uncertainty-aware methods \citep{c:55,c:54} achieved competitive RMSE of approximately $1.67$ mmol/L via closed-form neural dynamics or temporal--spectral fusion but showed limited clinical compliance with Zone A accuracy below $80\%$. Self-supervised pretraining approaches \citep{c:15} were suboptimal and did not effectively utilize unlabeled PPG signals in real-world data streams.
	
	Our paradigm addresses these challenges jointly. Dynamic incremental learning updates the model from an evolving stream with shifting distributions, adapting to intra-individual drift and reducing reliance on static training data. The uncertainty-quantified proxy gradient bridging agent enhances learning efficiency by extracting informative representations under sparse annotation. Mutual entropy--optimized replay clustering improves adaptability across meta- and real-world scenarios by organizing distributionally heterogeneous samples into latent task partitions for replay-based continual updating. The comparative results in Table~\ref{tab:2}, together with the Clarke Error Grid in Fig.~\ref{fig13} and the aggregation-aware analyses in Table~\ref{tab:3} and Table~\ref{tab:slice_robustness}, provide complementary evidence for the robustness of the proposed framework under subject-independent evaluation.
	
	As listed in Table~\ref{tab:2}, handcrafted pipelines leveraging time-, frequency-, and entropy-domain descriptors with shallow regressors \citep{c:10,c:11,c:16} yield a best reproduced MAE/RMSE of $\ge 1.50/2.58$ mmol/L under the same fixed-configuration protocol. This gap suggests that adaptive feature learning, dynamic task organization, and continual updating are beneficial for this benchmark, although the comparison should be understood within the reproduced-default setting rather than as an exhaustive hyperparameter-optimized ranking of all possible implementations.
	
	From an error-grid perspective, \(97.69\%\) of the predictions fall within Clarke A+B zones (Fig.~\ref{fig13}), indicating that most estimates remain in clinically acceptable decision regions. We further examined the same held-out test-fold predictions using ISO 15197:2013 reference error bands as a retrospective measurement-quality analysis rather than as evidence of regulatory compliance. The detailed window-, acquisition-, and subject-level error-band statistics are reported in Section~\ref{sec:aggregation_slice_robustness}. Compared with reproduced SOTA methods, the proposed framework reduces MAE/RMSE by \(23\%\)--\(86\%\) while preserving clinically meaningful error-grid behavior. The fold-isolated evaluation protocol, aggregation-aware statistics, and glucose-range analyses provide a reproducible basis for subsequent benchmarking. Collectively, these results indicate that adaptive, label-efficient learning can improve non-invasive BG estimation under longitudinal distribution shift, while formal device-level compliance would require a dedicated prospective validation protocol.

	\subsection{Aggregation- and Slice-wise Robustness Analysis}
	\label{sec:aggregation_slice_robustness}
	
	To further examine whether the reported performance remains stable beyond aggregate window-level statistics, we reanalyzed the same frozen predictions from the subject-independent test folds without retraining the model. Since each 40-s PPG acquisition is divided into ten non-overlapping 4-s windows sharing the same reference capillary BG measurement, we first evaluated the predictions at multiple aggregation levels. As shown in Table~\ref{tab:3}, \(91.2\%\) of window-level predictions fell within the ISO 15197:2013 reference error bands. After averaging the valid 4-s predictions from the same 40-s acquisition before evaluation, the within-band percentage increased to \(93.8\%\) with a 95\% confidence interval of \(92.0\%\)--\(95.4\%\). Subject-macro analysis yielded a comparable within-band percentage of \(93.6\%\) with a 95\% confidence interval of \(90.4\%\)--\(96.2\%\). These results indicate that the favorable error distribution is not explained only by treating adjacent windows as independent samples or by a small number of frequently sampled subjects. The ISO-related results are therefore reported as retrospective measurement-quality analyses rather than as evidence of formal regulatory compliance.
	
	\begin{table}[pos=t]
	\centering
	\normalfont\footnotesize
	\setlength{\tabcolsep}{4pt}
	\renewcommand{\arraystretch}{1.12}
	\caption{ISO-related error-band analysis at different aggregation levels. The ISO 15197:2013 error bands are used as reference measurement-quality bands rather than as evidence of regulatory compliance.}
	\label{tab:3}
	\begin{threeparttable}
		\begin{tabular*}{\linewidth}{@{\extracolsep{\fill}}llccc@{}}
			\toprule
			Analysis level
			& Unit of analysis
			& MAE \(\downarrow\)
			& Within band (\%) \(\uparrow\)
			& 95\% CI \\
			\midrule
			Window level
			& 4-s PPG window
			& 0.64
			& 91.2
			& 90.4--91.9 \\
			Acquisition level
			& 40-s PPG acquisition
			& 0.60
			& 93.8
			& 92.0--95.4 \\
			Subject-macro level
			& Subject-wise macro average
			& 0.65
			& 93.6
			& 90.4--96.2 \\
			\bottomrule
		\end{tabular*}
		\begin{tablenotes}[flushleft]
			\normalfont\scriptsize
			\item[*] \raggedright Note. The ISO-related reference bands were defined as \(\pm0.83\) mmol/L for reference BG below \(5.55\) mmol/L and \(\pm15\%\) for reference BG at or above \(5.55\) mmol/L. Acquisition-level metrics were computed after averaging the valid 4-s window predictions belonging to the same 40-s acquisition. Subject-macro metrics were computed by first obtaining subject-specific acquisition-level metrics and then averaging across subjects. Confidence intervals were estimated by non-parametric bootstrap resampling over the corresponding aggregation unit.
		\end{tablenotes}
	\end{threeparttable}
\end{table}
	
	We then summarized performance by acquisition unit, subject, glucose range, and acquisition period. Table~\ref{tab:slice_robustness} reports the corresponding robustness analysis. Acquisition-level aggregation slightly reduced MAE compared with the original window-level evaluation, which is expected because adjacent 4-s windows from the same 40-s acquisition capture short-term waveform fluctuations around the same reference BG value. The subject-macro results remained close to the overall performance, indicating that the reported accuracy is not dominated by a small subset of frequently sampled subjects. Across glucose ranges, the normal-to-moderately elevated range showed the most stable behavior, whereas the high-glucose range remained more challenging, with larger RMSE and lower ISO-related within-band percentage. This pattern is consistent with the smaller number of high-glucose samples and the stronger physiological variability that can occur under hyperglycemic or clinically unstable conditions. The temporal slices also showed limited performance variation across the acquisition period. Although the late acquisition period exhibited a slightly higher RMSE, the MAE and Clarke A+B percentages remained close to the overall test-fold performance. These findings support the interpretation that the proposed dynamic incremental learning framework mitigates longitudinal distribution shift under the subject-independent protocol, while also identifying high-glucose and clinically unstable regimes as important targets for future data enrichment.
	
	\begin{table}[pos=t] \centering \normalfont\scriptsize \setlength{\tabcolsep}{3pt} \renewcommand{\arraystretch}{1.10} \caption{Aggregation- and slice-wise robustness analysis using the same frozen subject-independent test-fold predictions.} \label{tab:slice_robustness} \begin{tabular*}{\linewidth}{@{\extracolsep{\fill}}llcccc@{}} \toprule Slice type & Subset & MAE $\downarrow$ & RMSE $\downarrow$ & Clarke A+B $\uparrow$ & ISO-related within band $\uparrow$ \\ \midrule Aggregation & Window level & 0.64 & 1.29 & 97.69 & 91.2 \\ Aggregation & 40-s acquisition level & 0.60 & 1.16 & 98.10 & 93.8 \\ Aggregation & Subject-macro level & 0.65 & 1.25 & 97.40 & 93.6 \\ \midrule Glucose range & $y<5.55$ mmol/L & 0.52 & 0.78 & 98.60 & 86.9 \\ Glucose range & $5.55\le y<10.0$ mmol/L & 0.57 & 1.02 & 98.30 & 93.9 \\ Glucose range & $y\ge10.0$ mmol/L & 1.18 & 2.34 & 94.20 & 81.7 \\ \midrule Temporal period & Early acquisition & 0.63 & 1.23 & 98.10 & 91.8 \\ Temporal period & Middle acquisition & 0.65 & 1.31 & 97.50 & 90.9 \\ Temporal period & Late acquisition & 0.66 & 1.36 & 97.10 & 90.3 \\ \bottomrule \end{tabular*} \end{table}
	
The vertical spread observed near reference BG values around 16~mmol/L in Fig.~\ref{fig13} can also be interpreted from the aggregation structure of the benchmark. Each 40-s acquisition is paired with a single reference capillary BG measurement but contributes multiple non-overlapping 4-s PPG windows. Several window-level predictions can therefore share the same reference value and appear as a vertical column in the Clarke Error Grid. This pattern is more visible in the high-glucose range, where samples are less frequent and may be more affected by peripheral perfusion variation, vascular tone changes, acute physiological state, and reference-measurement noise. Consistent with this interpretation, the glucose-range analysis in Table~\ref{tab:slice_robustness} shows that the high-glucose subgroup (\(y\ge10.0\) mmol/L) has a higher RMSE of \(2.34\) mmol/L and a lower ISO-related within-band percentage of \(81.7\%\), compared with \(1.02\) mmol/L and \(93.9\%\) in the \(5.55\le y<10.0\) mmol/L range. This region is therefore interpreted as a failure-prone subgroup requiring further data enrichment rather than as evidence of uniform predictive stability across the entire glucose spectrum.

	\subsection{Ablation Study}
	\subsubsection{Performance Comparison of Proposed Methods}
\begin{figure}[pos=t]
		\centering
		\includegraphics[width=1\linewidth]{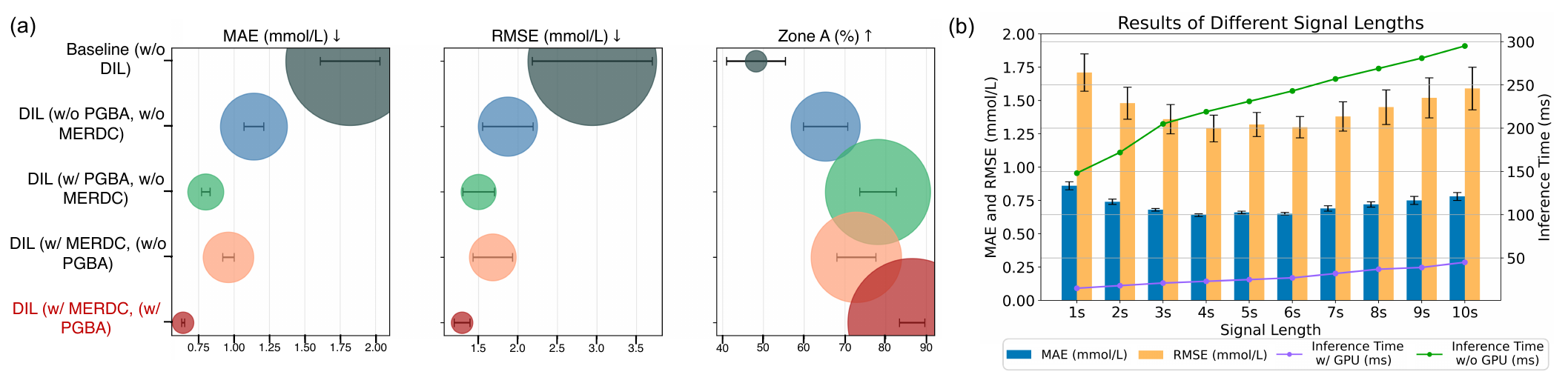}
		\caption{Ablation study of the proposed method, where (a) presents the baseline component analysis and (b) shows the integrated results combining the ablation study outcomes with the evaluation under varying signal lengths.}
		\label{fig12}
		\vspace{-1em}
	\end{figure}

	The ablation results in Fig.~\ref{fig12}(a) demonstrate the incremental contributions of the proposed components, namely Dynamic Incremental Learning (DIL), the Uncertainty-Quantified Proxy Gradient Bridging Agent (PGBA), and Mutual Entropy-Optimized Replay-Based Dynamic Clustering (MERDC), to the backbone model. The figure comprises three panels: the left shows Mean Absolute Error (MAE, mmol/L), the middle presents Root Mean Square Error (RMSE, mmol/L), and the right depicts the percentage of predictions within Zone A of the Clarke Error Grid. Each data point is a colored circle corresponding to a specific configuration, positioned at the mean metric value. Horizontal bars within each circle represent error bars from 5-fold cross-validation, denoting the standard deviation across folds and thus reflecting variability and robustness under different partitions.

	The component-level ablation in Fig.~\ref{fig12}(a) and Table~\ref{tab:component_contribution} summarizes the marginal role of each module. The backbone-only model provides the non-incremental reference and shows substantial error and fold variability, indicating that a static estimator is insufficient for the non-stationary PPG--glucose mapping. Adding DIL reduces MAE from \(1.82\pm0.21\) to \(1.14\pm0.07\) mmol/L and improves Zone A from \(48.16\pm7.21\%\) to \(65.23\pm5.45\%\), showing that bounded replay and gradient-projection constraints provide the primary stability--plasticity benefit. Adding PGBA to DIL further reduces MAE to \(0.80\pm0.03\) mmol/L, supporting the value of exploiting unlabeled PPG windows through proxy-gradient learning. Adding MERDC to DIL yields an MAE of \(0.96\pm0.04\) mmol/L, indicating that ME2AC-driven task organization and replay construction improve adaptation even without PGBA. The full DIL+PGBA+MERDC configuration achieves the best overall performance, with MAE \(0.64\pm0.01\) mmol/L, RMSE \(1.29\pm0.10\) mmol/L, and Zone A \(86.48\pm3.10\%\). This pattern suggests that the three modules are complementary rather than redundant: DIL provides the continual-learning backbone, PGBA improves label efficiency through unlabeled streams, and MERDC supplies the latent task structure used for replay and incremental updating.
	
	\begin{table}[pos=t]
		\centering
		\normalfont\footnotesize
		\setlength{\tabcolsep}{3.5pt}
		\renewcommand{\arraystretch}{1.12}
		\caption{Component-level contribution analysis of the proposed framework under subject-independent five-fold validation.}
		\label{tab:component_contribution}
		\begin{tabular*}{\linewidth}{@{\extracolsep{\fill}}lccccp{3.1cm}@{}}
			\toprule
			Configuration
			& MAE $\downarrow$
			& RMSE $\downarrow$
			& Zone A $\uparrow$
			& MAE reduction
			& Main tested contribution \\
			\midrule
			Backbone only
			& \(1.82\pm0.21\)
			& \(2.94\pm0.76\)
			& \(48.16\pm7.21\)
			& --
			& Static non-incremental reference \\
			DIL
			& \(1.14\pm0.07\)
			& \(1.87\pm0.32\)
			& \(65.23\pm5.45\)
			& \(37.4\%\) vs. backbone
			& Continual replay and gradient projection \\
			DIL+PGBA
			& \(0.80\pm0.03\)
			& \(1.50\pm0.20\)
			& \(78.10\pm4.50\)
			& \(29.8\%\) vs. DIL
			& Unlabeled-stream proxy-gradient learning \\
			DIL+MERDC
			& \(0.96\pm0.04\)
			& \(1.68\pm0.25\)
			& \(72.80\pm4.80\)
			& \(15.8\%\) vs. DIL
			& ME2AC-guided task discovery and replay organization \\
			DIL+PGBA+MERDC
			& \(\mathbf{0.64\pm0.01}\)
			& \(\mathbf{1.29\pm0.10}\)
			& \(\mathbf{86.48\pm3.10}\)
			& \(20.0\%\) vs. DIL+PGBA
			& Integrated label-efficient dynamic adaptation \\
			\bottomrule
		\end{tabular*}
	\end{table}

	\subsubsection{PGBA Reliability Audit} \label{sec:pgba_reliability} To examine whether PGBA provides useful proxy-gradient directions, we conducted a reliability audit on labeled mini-batches from the training folds. These audit samples are suitable for this analysis because both the true supervised output-space gradient \(q_i=\partial\ell/\partial\mathbf{J}(\mathbf{x}_i)\) and the PGBA proxy gradient \(\bar{g}_i\) are available. We report three quantities: the cosine similarity between \(q_i\) and \(\bar{g}_i\), the positive-alignment ratio, and the relative bridge-loss reduction after the one-step proxy update: \[ \Delta_{\mathrm{bridge}} = \frac{ \ell(\mathbf{J}_i,y_i)-\ell(\mathbf{J}_i-\eta\bar{g}_i,y_i) }{ \ell(\mathbf{J}_i,y_i)+\xi }. \] A positive \(\Delta_{\mathrm{bridge}}\) indicates that the proxy direction locally reduces the supervised loss on labeled audit samples. \begin{table}[pos=htbp] \centering \normalfont\footnotesize \setlength{\tabcolsep}{5pt} \caption{Training-fold reliability audit of PGBA under subject-independent five-fold validation.} \label{tab:pgba_reliability} \begin{tabular*}{\linewidth}{@{\extracolsep{\fill}}lccccc@{}} \toprule Configuration & Cosine $\uparrow$ & Positive alignment (\%) $\uparrow$ & \(\Delta_{\mathrm{bridge}}\) (\%) $\uparrow$ & MAE $\downarrow$ & RMSE $\downarrow$ \\ \midrule DIL+PGBA & 0.38 $\pm$ 0.06 & 82.4 $\pm$ 4.8 & 3.6 $\pm$ 1.1 & 0.80 $\pm$ 0.03 & 1.50 $\pm$ 0.20 \\ DIL+PGBA+MERDC & 0.45 $\pm$ 0.05 & 87.9 $\pm$ 3.9 & 5.1 $\pm$ 1.2 & 0.64 $\pm$ 0.01 & 1.29 $\pm$ 0.10 \\ \bottomrule \end{tabular*} \end{table} As shown in Table~\ref{tab:pgba_reliability}, the proxy gradients learned by PGBA show positive average alignment with the supervised output-space gradients on labeled training-fold audit samples. Most audited samples have positive proxy--supervised alignment, and the bridge-loss reduction is positive for both configurations. These results support the interpretation that PGBA does not generate arbitrary update directions. Instead, the bridge-loss training encourages proxy gradients that locally reduce the supervised loss. The full DIL+PGBA+MERDC configuration further improves alignment and bridge-loss reduction, suggesting that ME2AC-derived task organization provides a more stable representation space for proxy-gradient learning.
	\subsubsection{Performance Comparison of Different Input Signal Lengths}
	In Fig.~\ref{fig12}(b), we examine the impact of different sampling durations on blood glucose estimation. A 4-second signal yields the lowest MAE and RMSE ($0.64$ and $1.29$ mmol/L, respectively), indicating that this duration captures salient features while limiting noise accumulation. At 6 seconds, performance declines slightly (MAE $=0.65$, RMSE $=1.30$), likely due to redundancy introducing interference.
	
	For 1--2 second signals, MAE is higher ($>0.74$ mmol/L), reflecting the challenge of modeling dynamic physiological processes with very short sequences, consistent with the time-dependent nature of glucose metabolism. Beyond 7 seconds, errors continue to rise, primarily because longer segments increase exposure to irrelevant temporal noise.
	
	Accordingly, selecting a 4-second window strikes a favorable balance between accuracy (MAE $<0.65$ mmol/L) and real-time performance (GPU inference time $<25$ ms), aligning with the responsiveness needs of high-frequency wearable measurement systems.
	
	These results suggest the presence of a ``critical time window'' for BG estimation. Future work could integrate frequency-domain analysis to further probe the relationship between window length and physiological rhythms, with the goal of improving model generalization.
	
	\begin{table}[pos=htbp]
		\centering
		\normalfont\footnotesize
		\setlength{\tabcolsep}{3pt}
		\caption{Performance comparison of backbone architectures for PPG-based glucose estimation.}
		\begin{threeparttable}
			\begin{tabular*}{\linewidth}{@{\extracolsep{\fill}}lccccc@{}}
				\toprule
				\multirow{2}{*}{Backbone} & MAE & \multirow{2}{*}{PCC $\uparrow$} & Zone A & Params & Inf. Time  \\
				& (mmol/L) $\downarrow$ & & (\%) $\uparrow$  & (M) $\downarrow$ & (ms) $\downarrow$\\
				\midrule
				ResNet (2016) & 0.71 $\pm$0.03 & 0.72 $\pm$0.04 & 78.32 $\pm$2.45 & 13.20 & 42 \\
				LSTM-FCN (2017) & 0.68 $\pm$0.02 & 0.75 $\pm$0.03 & 80.15 $\pm$3.12 & 6.92 & 58 \\
				GRU-FCN (2018) & 0.67 $\pm$0.02 & 0.77 $\pm$0.03 & 81.04 $\pm$2.89 & 8.71 & 49 \\
				InceptionTime (2019) & 0.64 $\pm$0.01 & 0.88 $\pm$0.01 & 86.48 $\pm$3.10 & 6.84 & 23 \\
				OmniScale (2020) & 0.63 $\pm$0.01 & 0.82 $\pm$0.02 & 84.25 $\pm$2.78 & 12.82 & 31 \\
				gMLP (2021) & 0.62 $\pm$0.01 & 0.83 $\pm$0.02 & 83.91 $\pm$2.95 & 28.41 & 65 \\
				TSPerceiver (2022) & 0.66 $\pm$0.02 & 0.81 $\pm$0.02 & 83.12 $\pm$3.14 & 41.79 & 89 \\
				PatchTST (2023) & 0.64 $\pm$0.01 & 0.84 $\pm$0.02 & 85.02 $\pm$2.67 & 29.63 & 28 \\
				StarNet (2024) & 0.63 $\pm$0.01 & 0.83 $\pm$0.01 & 84.73 $\pm$2.84 & 8.76 & 30 \\
				TFDNet (2025) & 0.62 $\pm$0.01 & 0.85 $\pm$0.01 & 85.91 $\pm$2.55 & 9.35 & 25 \\
				\bottomrule
			\end{tabular*}
			\begin{tablenotes}
				\normalfont\scriptsize
				\item[*] PCC: Pearson Correlation Coefficient, Params: Trainable parameters (millions), Inf. Time: Single-sample inference time on GPU.
			\end{tablenotes}
		\end{threeparttable}
		\label{tab:5}
	\end{table}
	\subsubsection{Performance Comparison of Different Backbones}
	We follow the backbone network adopted in the literature \citep{c:15} and reference recent architectures such as StarNet \citep{c:56} and TFDNet \citep{c:57}. Within the proposed paradigm, we evaluate multiple backbones to inform model selection and demonstrate framework versatility. As shown in Table \ref{tab:5}, InceptionTime offers the best balance between estimation accuracy and computational efficiency. Although TFDNet (MAE $=0.62$) and gMLP (MAE $=0.62$) perform slightly better on absolute error, InceptionTime achieves the most balanced overall profile in CEG-based reliability, physiological consistency, and on-device efficiency. It attains the highest Zone A percentage ($86.48\%$ vs. TFDNet’s $85.91\%$), indicating stronger performance within clinically acceptable ranges. In addition, InceptionTime exhibits the strongest agreement with reference measurements (PCC $=0.88$ vs. PatchTST’s $0.84$), suggesting better preservation of blood glucose trends. It also uses $26.8\%$ fewer parameters than TFDNet (6.84M vs. 9.35M) and achieves the fastest inference (23 ms), which is advantageous for future wearable implementations.
	
	InceptionTime’s multi-scale convolutional blocks enable end-to-end learning of high-frequency intra-beat patterns (e.g., upstroke, systolic peak, dicrotic notch, diastolic decay) and low-frequency inter-beat trends (e.g., pulse rate variability and amplitude baseline fluctuations related to vascular compliance) directly from raw windows, avoiding hand-engineered feature extractors. Its hierarchical feature fusion, compared with TFDNet’s frequency-domain decomposition, is well suited to capturing short-term PPG changes (within 4 s) alongside longer-term BG dynamics. Furthermore, its parameter sharing reduces model size by $58\%$ compared with the standard Inception architecture while maintaining temporal modeling capacity. Although newer architectures such as TFDNet excel on specific metrics, their higher complexity (e.g., 41.79M parameters in TSPerceiver) constrains practical deployment. Overall, model choice for continuous BG estimation should balance clinical effectiveness with operational constraints.
	
	\begin{table}[pos=htbp]
		\centering
		\normalfont\footnotesize
		\setlength{\tabcolsep}{3pt}
		\caption{Performance comparison of clustering operators used within dynamic incremental learning (DIL).}
		\begin{threeparttable}
			\begin{tabular*}{\linewidth}{@{\extracolsep{\fill}}lccccc@{}}
				\toprule
				\makecell[l]{Method} 
				& \makecell[c]{MAE\\(mmol/L) $\downarrow$} 
				& \makecell[c]{Runtime\\(ms) $\downarrow$} 
				& \makecell[c]{Silhouette\\Coefficient $\uparrow$} 
				& \makecell[c]{Davies-Bouldin\\Index $\downarrow$} 
				& \makecell[c]{Feature\\Purity $\uparrow$} \\
				\midrule
				DBSCAN & 0.89 & \textbf{145} & 0.51 & 1.89 & 0.68 \\
				HDBSCAN & 0.93 & 231 & 0.63 & 1.42 & 0.72 \\
				OPTICS & 1.05 & 317 & 0.58 & 1.65 & 0.65 \\
				Mean Shift & 1.12 & 489 & 0.47 & 2.15 & 0.61 \\
				\textbf{ME2AC (Ours)} & \textbf{0.64} & 179 & \textbf{0.71} & \textbf{0.93} & \textbf{0.84} \\
				\bottomrule
			\end{tabular*}
		\end{threeparttable}
		\label{tab:6}
	\end{table}
	\subsubsection{Performance Comparison of Different Unsupervised Clustering Methods}
	We adopt commonly used metrics for unsupervised clustering to fairly evaluate ME2AC, the clustering operator embedded in MERDC, as shown in Table \ref{tab:6}. ME2AC outperforms baseline methods in terms of Mean Absolute Error (MAE), achieving $0.64$ mmol/L, a $42.9\%$ improvement over standard DBSCAN. This gain is attributed to the dynamic neighborhood radius calculation (Eq. \myref{eq:9}), which adapts to local density variations in standardized PPG-window neighborhoods and effectively addresses the fixed $\varepsilon$ limitation of traditional DBSCAN.
	
	Although ME2AC has a longer runtime than DBSCAN due to the inclusion of entropy computation (Eq. \myref{eq:10}) and histogram-based mutual-information analysis (Eq. \myref{eq:11}), it remains computationally efficient. The median-based kNN distance calculation reduces unnecessary neighborhood expansion compared with the hierarchical approach of HDBSCAN. ME2AC attains the highest silhouette coefficient of 0.71 and the lowest Davies--Bouldin index of 0.93, supporting its dual optimization strategy of minimizing intra-cluster entropy and maximizing inter-cluster mutual information. This balance effectively prevents over-segmentation, which is a common issue in OPTICS, and under-clustering, which is prevalent in Mean Shift.
	
	Feature purity improves by $24.6\%$ (0.84 vs. DBSCAN’s 0.68), a direct result of mutual information–based cluster merging (Eq. \myref{eq:11}). This improvement indicates that the operator better preserves physiological relevance during PPG-window task discovery. The linear time complexity in the neighborhood expansion stage makes ME2AC suitable for large-scale Dynamic Incremental Learning (DIL) applications and helps explain the downstream gains of the higher-level MERDC module. The mutual information threshold $\tau_{\text{MI}}$ serves as an automatic regularization parameter, preventing cluster inflation during incremental updates. 
	The innovation of ME2AC lies in its dual information-theoretic optimization: entropy minimization promotes cluster compactness without relying on outlier-sensitive means, while histogram-based mutual-information merging preserves clinically pertinent similarity patterns during cluster merging. This combination addresses the fundamental trade-off between cluster purity and preservation of physiological relevance in dynamic PPG analysis and forms the clustering foundation of MERDC.

\subsubsection{Sensitivity of ME2AC Information-Estimation Parameters}
\label{sec:me2ac_sensitivity}

The preceding comparison shows that ME2AC provides better clustering quality and downstream prediction accuracy than conventional unsupervised clustering methods. We further examined whether this improvement depends on fragile choices of histogram or threshold parameters. Specifically, we varied the entropy-bin number $K_H$, the mutual-information-bin number $K_{MI}$, the entropy quantile $\alpha_H$, and the mutual-information quantile $\alpha_{MI}$ under the same subject-independent 5CV protocol. For all settings, standardization statistics, histogram ranges, equivalence bins, and thresholds were estimated only from the corresponding training folds.

\begin{table}[pos=t]
	\centering
	\caption{Sensitivity analysis of ME2AC information-estimation parameters under subject-independent 5-fold cross-validation. The ``Tasks across folds'' column reports the number of tasks discovered in each of the five training folds.}
	\label{tab:me2ac_sensitivity}
	\normalfont\scriptsize
	\setlength{\tabcolsep}{2.5pt}
	\begin{tabular*}{\linewidth}{@{\extracolsep{\fill}}llccccc@{}}
		\toprule
		Factor & Setting
		& \makecell[c]{Tasks\\across folds}
		& \makecell[c]{Silhouette\\$\uparrow$}
		& \makecell[c]{DB index\\$\downarrow$}
		& \makecell[c]{Feature\\purity $\uparrow$}
		& \makecell[c]{MAE\\(mmol/L) $\downarrow$} \\
		\midrule
		Default & $K_H=10$, $K_{MI}=10$, $\alpha_H=0.60$, $\alpha_{MI}=0.75$
		& 5/5/5/5/6 & 0.71 & 0.93 & 0.84 & 0.64 \\
		\midrule
		Entropy bins & $K_H=8$
		& 5/5/5/5/5 & 0.70 & 0.96 & 0.83 & 0.65 \\
		Entropy bins & $K_H=12$
		& 5/5/5/6/6 & 0.70 & 0.95 & 0.83 & 0.65 \\
		\midrule
		MI bins & $K_{MI}=8$
		& 5/5/5/5/6 & 0.70 & 0.96 & 0.83 & 0.65 \\
		MI bins & $K_{MI}=12$
		& 5/5/5/6/6 & 0.70 & 0.95 & 0.84 & 0.65 \\
		\midrule
		Entropy threshold & $\alpha_H=0.50$
		& 4/5/5/5/5 & 0.69 & 0.99 & 0.82 & 0.66 \\
		Entropy threshold & $\alpha_H=0.70$
		& 5/6/6/6/6 & 0.69 & 0.98 & 0.82 & 0.65 \\
		\midrule
		MI threshold & $\alpha_{MI}=0.65$
		& 4/4/5/5/5 & 0.68 & 1.01 & 0.81 & 0.66 \\
		MI threshold & $\alpha_{MI}=0.85$
		& 6/6/6/6/6 & 0.69 & 0.97 & 0.83 & 0.65 \\
		\bottomrule
	\end{tabular*}
\end{table}

As shown in Table~\ref{tab:me2ac_sensitivity}, ME2AC remains stable across reasonable variations of histogram and threshold parameters. The default configuration discovers five to six tasks across the five subject-independent folds and achieves the best or near-best clustering compactness and downstream prediction accuracy. Lower mutual-information quantiles tend to merge more preliminary clusters and produce fewer discovered tasks, whereas higher quantiles preserve more task partitions. Similarly, a looser entropy threshold increases the number of core points and slightly increases the number of discovered tasks. These trends are consistent with the intended roles of the entropy and mutual-information terms, suggesting that ME2AC is not driven by a fragile binning or threshold choice.

\subsubsection{Task-Structure Role of ME2AC in Continual Learning}
\label{sec:me2ac_task_structure}

The preceding clustering comparison and sensitivity analysis evaluate ME2AC from the perspective of clustering quality and parameter robustness. We further clarify the role of the discovered partitions in the continual-learning pipeline. The task indices produced by ME2AC are fold-local latent task partitions rather than predefined clinical categories or discretized glucose classes. Accordingly, their practical value lies not in assigning diagnostic labels to clusters, but in organizing heterogeneous training-fold samples into operational units for sequential learning, replay construction, and distribution-selective updating.

\begin{table}[pos=t]
	\centering
	\normalfont\footnotesize
	\setlength{\tabcolsep}{3pt}
	\renewcommand{\arraystretch}{1.18}
	\caption{Structural evidence and continual-learning role of ME2AC-discovered latent task partitions.}
	\label{tab:me2ac_structure_role}
	\begin{tabularx}{\linewidth}{
			@{}
			>{\raggedright\arraybackslash}p{0.20\linewidth}
			>{\raggedright\arraybackslash}p{0.23\linewidth}
			>{\raggedright\arraybackslash}p{0.17\linewidth}
			>{\raggedright\arraybackslash}X
			@{}
		}
		\toprule
		Aspect
		& Empirical observation
		& Source
		& Interpretation for continual learning \\
		\midrule
		Number of discovered tasks
		& \(5/5/5/5/6\)
		& Default 5CV setting
		& ME2AC identifies a small and stable set of latent task partitions, rather than producing excessive fragmented clusters. \\
		
		Cluster compactness
		& Silhouette \(=0.71\)
		& Clustering comparison
		& Compact within-task structure supports stable task-wise replay and gradient constraints. \\
		
		Inter-partition separation
		& Davies--Bouldin index \(=0.93\)
		& Clustering comparison
		& Lower overlap between partitions helps reduce interference during sequential task updates. \\
		
		Feature consistency
		& Feature purity \(=0.84\)
		& Clustering comparison
		& The partitions preserve coherent latent PPG representations and can therefore serve as operational tasks for DIL. \\
		
		Parameter robustness
		& MAE remains \(0.64\)--\(0.66\) mmol/L across tested settings
		& ME2AC sensitivity analysis
		& The downstream benefit is not driven by a fragile histogram or threshold choice. \\
		
		Task-count stability
		& \(4\)--\(6\) tasks across sensitivity settings
		& ME2AC sensitivity analysis
		& Moderate parameter changes adjust the granularity of task discovery without destabilizing the learning pipeline. \\
		\bottomrule
	\end{tabularx}
\end{table}
As summarized in Table~\ref{tab:me2ac_structure_role}, ME2AC produces a compact and stable latent task structure under the subject-independent protocol. This structure is used directly by MERDC and DIL in three ways. First, the discovered task set \(\mathcal{T}\) defines the incremental curriculum, so that the model is updated over distributionally organized partitions rather than over an arbitrary shuffled stream. Second, the number of tasks \(B\) determines the task-balanced replay allocation, preventing dominant routine patterns from occupying the entire memory buffer. Third, when real-scenario data are integrated, ME2AC allows newly emerging patterns to be compared with existing task partitions through entropy and mutual-information criteria before they are assimilated into the incremental update. In this sense, ME2AC contributes to continual learning not merely by improving a clustering score, but by providing the task structure through which replay, gradient projection, and proxy-gradient learning are coordinated.

	\subsubsection{Performance Comparison of Different Folds}
	Across all folds, the MAE remains relatively stable, indicating robustness to local data distributions and noise (Table \ref{tab:8}). This is important for medical applications, as it may reduce the risk of erroneous interpretation. The Zone A coverage of $86.48\% \pm 3.10$ highlights strong performance within the clinically acceptable error range of $\le 20\%$. However, the relatively large standard deviation reflects fold-specific bias, with Fold 2 achieving $83.60\%$ and Fold 5 achieving $91.06\%$, suggesting imbalanced data distributions. To further improve generalization, stratified sampling or targeted data augmentation could be considered.

	\begin{table}[t]
		\centering
		\normalfont\scriptsize
		\setlength{\tabcolsep}{2.8pt}
		\renewcommand{\arraystretch}{1.12}
		\caption{Fold-wise point-estimation and uncertainty-calibration results under subject-independent 5-fold cross-validation.}
		\label{tab:8}
		\begin{threeparttable}
			\begin{tabular*}{\linewidth}{@{\extracolsep{\fill}}lcccccccc@{}}
				\toprule
				& \multicolumn{2}{c}{Point estimation}
				& \multicolumn{3}{c}{Clarke Error Grid}
				& \multicolumn{3}{c}{Uncertainty evaluation} \\
				\cmidrule(lr){2-3}
				\cmidrule(lr){4-6}
				\cmidrule(lr){7-9}
				Fold
				& MAE (mmol/L)
				& RMSE (mmol/L)
				& Zone A (\%)
				& Zone B (\%)
				& A+B (\%)
				& NLL
				& 95\% PICP (\%)
				& Gap$_{95}$ (p.p.) \\
				\midrule
				1 & 0.64 & 1.16 & 85.12 & 13.93 & 99.02 & 1.53 & 94.3 & 0.7 \\
				2 & 0.65 & 1.41 & 83.60 & 12.47 & 96.07 & 1.65 & 94.8 & 0.2 \\
				3 & 0.65 & 1.29 & 84.79 & 11.05 & 95.84 & 1.69 & 95.0 & 0.0 \\
				4 & 0.64 & 1.34 & 87.51 & 10.79 & 98.30 & 1.74 & 93.8 & 1.2 \\
				5 & 0.63 & 1.25 & 91.06 & 8.20  & 99.26 & 1.63 & 93.9 & 1.1 \\
				\midrule
				Mean $\pm$ SD
				& $0.64\pm0.01$
				& $1.29\pm0.10$
				& $86.48\pm3.10$
				& $11.29\pm2.13$
				& $97.70\pm1.63$
				& $1.65\pm0.08$
				& $94.4\pm0.6$
				& $0.7\pm0.5$ \\
				\bottomrule
			\end{tabular*}
			\begin{tablenotes}[flushleft]
				\normalfont\scriptsize
				\item[*] \raggedright \textit{Note.} NLL denotes Gaussian negative log-likelihood. PICP denotes prediction interval coverage probability, and Gap$_{95}$ denotes the absolute deviation from the nominal 95\% coverage, reported in percentage points. Lower MAE, RMSE, NLL, and Gap$_{95}$ indicate better performance. Zone B is reported for completeness, while higher Zone A and A+B indicate better clinical agreement. PICP is expected to be close to the nominal 95\% level.
			\end{tablenotes}
		\end{threeparttable}
	\end{table}
		\begin{figure}[pos=t]
		\centering
		\includegraphics[width=0.45\linewidth]{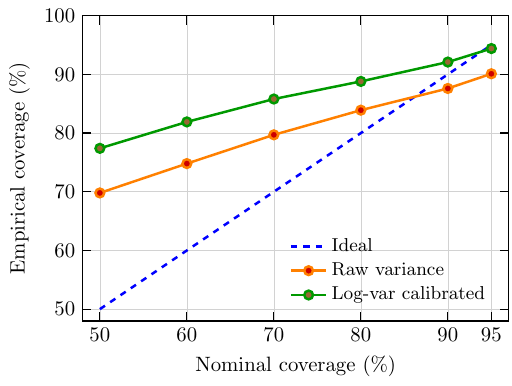}
		\caption{Uncertainty calibration curve under subject-independent 5-fold cross-validation. The dashed diagonal represents perfect agreement between nominal and empirical coverage. Compared with the raw variance output, log-variance calibration improves empirical coverage in the high-confidence range.}
		\label{fig:uncertainty_calibration}
	\end{figure}
	Table~\ref{tab:8} shows that the proposed method maintains stable point-estimation performance across the five subject-independent folds, with an average MAE of \(0.64\pm0.01\) mmol/L and RMSE of \(1.29\pm0.10\) mmol/L. The uncertainty evaluation further shows that the calibrated variance provides predictive intervals close to the nominal 95\% coverage, with a mean PICP of \(94.4\pm0.6\%\) and a small coverage gap of \(0.7\pm0.5\) percentage points. These results suggest that the auxiliary variance branch can serve as a reliability indicator after likelihood-based calibration, while leaving the point predictions used for MAE, RMSE, and Clarke Error Grid analysis unchanged.
	
	Figure~\ref{fig:uncertainty_calibration} further visualizes the empirical coverage of the uncertainty estimates across different nominal coverage levels. Compared with the raw variance output, the log-variance calibrated curve better approaches the nominal coverage in the high-confidence range, especially at the 90--95\% levels used in the main uncertainty evaluation. The lower nominal levels remain conservative, indicating that the calibrated intervals tend to avoid under-coverage rather than provide overly narrow uncertainty bands.

	\section{Discussion}
	\label{sec:discussions}
	\subsection{Long-Term Validation and Translational Relevance}
	A central question for non-invasive glucose estimation research is whether a model trained on encounter-level data can retain stable behavior over longer horizons in less controlled settings \citep{c:2, c:18}. While our main benchmark dataset provides robust validation across a large population, it consists of snapshot measurements. To examine longitudinal behavior beyond the primary benchmark, we conducted a pilot study over an extended observation period.
	
\begin{figure}[pos=t]
		\centering
		\includegraphics[width=0.94\linewidth]{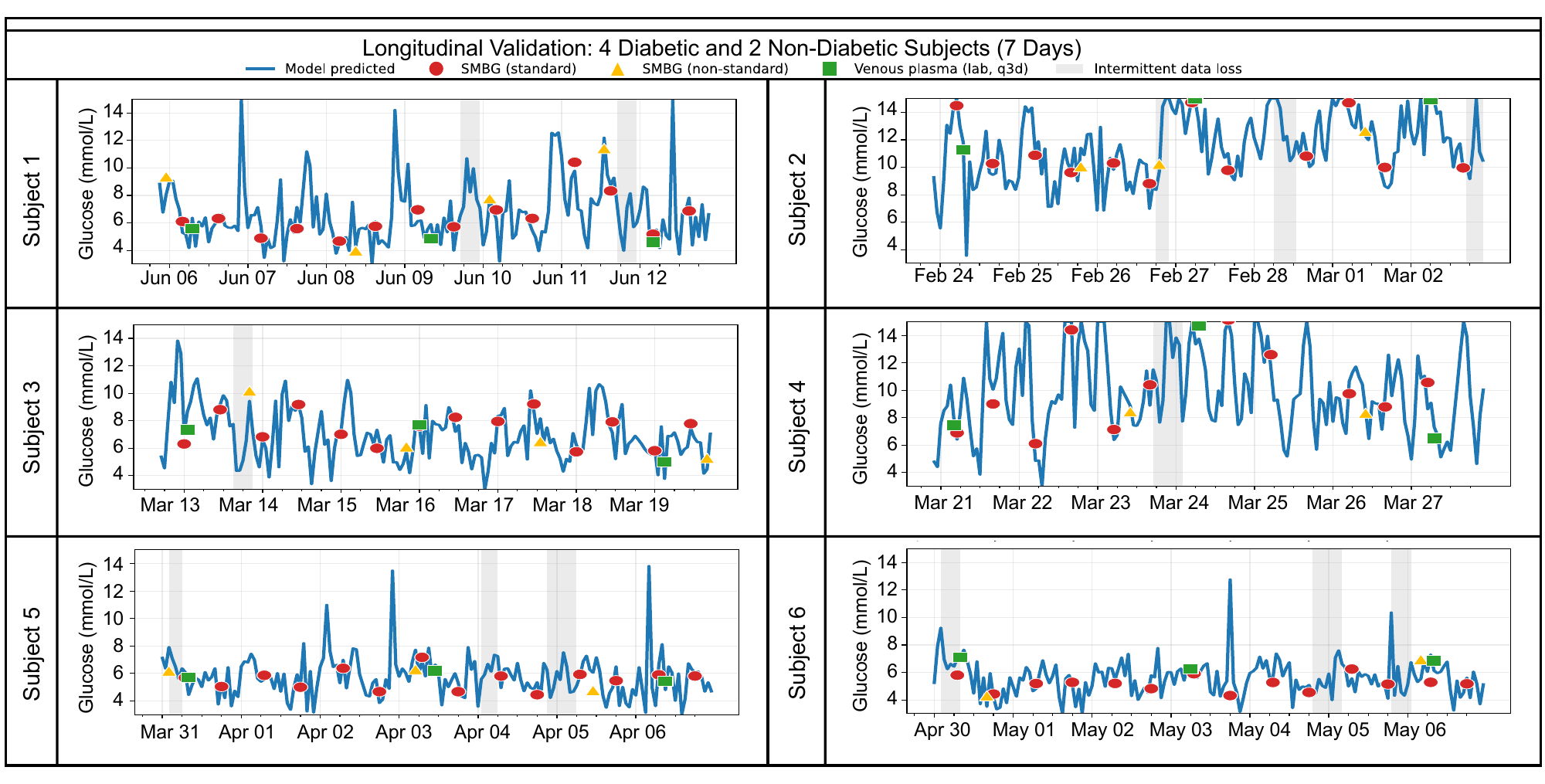}
		\caption{Seven-day pilot longitudinal validation across six participants (four with diabetes and two non-diabetic). Our model’s continuous PPG-based glucose prediction (solid lines) is compared with intermittent fingerstick SMBG, where circles represent standard fasting and postprandial measurements, and triangles indicate random SMBG tests, alongside laboratory venous glucose collected every three days (squares). Shaded areas denote periods of missing data (linearly interpolated) due to daily activities such as bathing or recharging.}
		\label{fig:long_term_validation}
		\vspace{-2em}
	\end{figure}

	As illustrated in Fig.~\ref{fig:long_term_validation}, we evaluated the system in a seven-day pilot feasibility study on six participants with varying health statuses, comprising four with diabetes and two non-diabetic individuals. Data were collected hourly using an alternate wrist-based PPG device for ambulatory acquisition. The figure presents the longitudinal results for all participants, comparing high-frequency glucose estimates from our DIL-based model with two reference sources: intermittent self-monitored blood glucose (SMBG) via fingerstick, where circles denote standard morning fasting and after-dinner postprandial measurements and triangles indicate additional random SMBG tests, as well as laboratory-analyzed venous plasma glucose collected every three days (squares). The plots show that our model tracks diurnal and postprandial glucose fluctuations and remains broadly consistent with the available references. In addition, the framework exhibits resilience to data gaps, which are common in real-world use due to activities such as bathing or battery recharging. Predictions remain stable upon data resumption, suggesting robust learned physiological representations. This pilot study therefore provides research-stage evidence that the approach can support longitudinal trend tracking between sparse reference measurements, while broader multi-center and prospective studies remain necessary to characterize translational performance more fully.
	
	\subsection{Catastrophic Forgetting}
\begin{figure}[pos=t]
		\centering
		\includegraphics[width=0.96\linewidth]{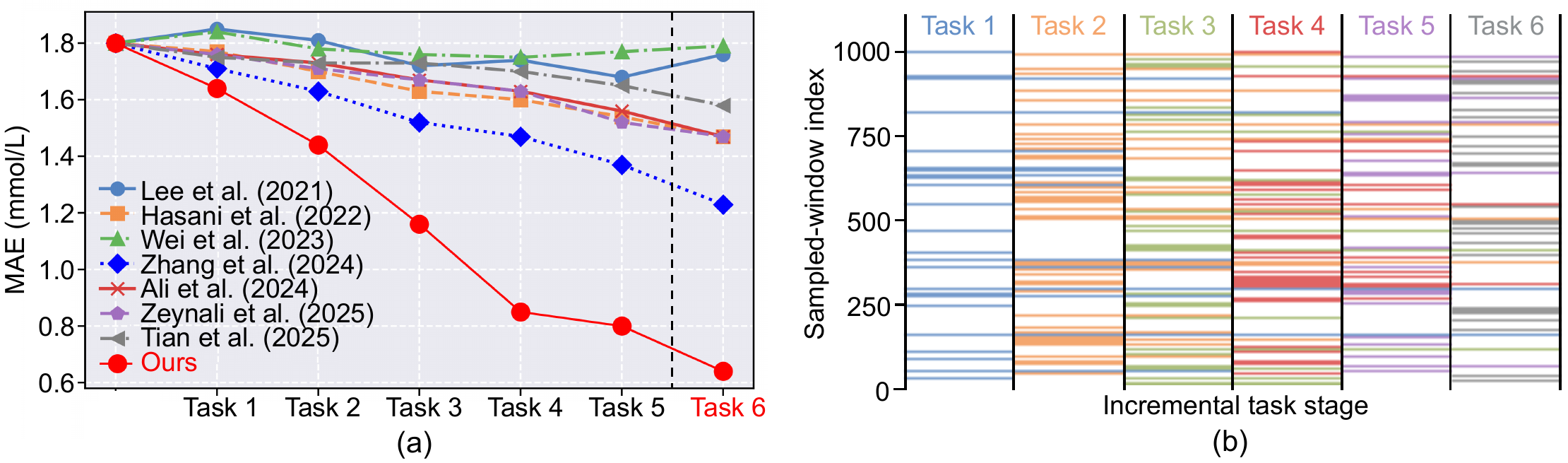}
		\caption{Illustration of dynamic adaptation under a non-stationary physiological data stream. (a) Method-wise MAE trajectories across incremental stages. The x-axis denotes the sequential task stage from Task~1 to Task~6, and the colored curves denote comparison methods rather than task identities. Task~6 corresponds to the newly instantiated incremental stage after releasing the unseen reservoir. (b) Illustrative participation pattern of a randomly sampled subset of 1,000 windows. The x-axis denotes the incremental task partition, the y-axis denotes the sampled-window index used for visualization, and colors denote ME2AC-discovered latent task partitions. This panel illustrates staged task participation and replay retention rather than a quantitative physiological axis.}
		\label{fig4}
		\vspace{-2em}
	\end{figure}

	As shown in Fig.~\ref{fig4}, non-stationary physiological acquisition (intra-individual regulatory drift, inter-subject heterogeneity, and device/environmental variability) induces a sequence of latent shifts in the joint PPG–glucose manifold that legacy pipelines handle poorly. Conventional strategies either fully retrain on the accumulated union of historical and newly collected data, incurring prohibitive computational and memory cost and aggravating catastrophic forgetting as documented in Table~\ref{tab:2}, or attempt superficial partial updates that lack principled mechanisms for identifying genuinely novel structure. The component analysis in Section~4.4 indicates that this behavior arises from the interaction between continual replay, proxy-gradient learning, and ME2AC-guided task organization, rather than from a single isolated module.
	
	 Self-supervised pretraining \citep{c:15} affords only limited resilience because it still presumes subsequent static fine-tuning. To emulate an open longitudinal-use scenario rather than a closed benchmark, we implement a single 5-fold partition and reserve one fold (20\%) from the outset as an unseen prospective reservoir used later to simulate real-world inflow, while another fold (20\%) is held out throughout as a dedicated test set. The remaining three folds (60\%) constitute the developmental stream.
	
	This 60\% developmental stream is unsupervisedly decomposed by MERDC into five distributionally heterogeneous latent task partitions, denoted Tasks~1--5. These task indices are fold-local operational partitions rather than predefined clinical categories or discretized glucose classes. In Fig.~\ref{fig4}(a), the horizontal axis represents the incremental stages from Task~1 to Task~6, whereas the colored curves denote comparison methods. Each latent task partition is characterized by information-theoretic profiles, including local entropy, mutual-information overlap, and inter-partition divergence, and is subsequently used to define the incremental curriculum and task-balanced replay memory in DIL. The resulting sequence is processed using bounded episodic memory and gradient-projection constraints that enforce non-interference between tasks, yielding a consolidated representation after Task~5 without any exposure to the withheld 40\%.
	
	Following convergence on Task 5, the unseen reservoir is released to emulate continuous acquisition. Instead of ingesting the entire pool or retraining on the full historical union, MERDC is reapplied to the arriving reservoir jointly with compact memory exemplars to test for distributional novelty. 
	
	Only the subset whose information signature exhibits insufficient mutual-information overlap and reduced redundancy relative to existing task manifolds is instantiated as a new incremental unit, denoted Task~6 on the horizontal task axis of Fig.~\ref{fig4}(a), while the remainder does not trigger redundant updates.
	
	Earlier tasks are retained solely through their episodic representatives, and raw historical samples are not reloaded. The model is updated exclusively on Task~6 plus replay, so computational and memory cost scale with marginal novelty rather than cumulative volume. This selective assimilation mitigates catastrophic forgetting by ensuring that previous loss constraints remain satisfied, avoids distributional dilution that would arise from indiscriminate mixing, and enforces stable manifold preservation through projection feasibility. Fig.~\ref{fig4}(b) visualizes the participation pattern using a randomly sampled illustrative subset of 1,000 windows for clarity. The vertical axis indexes sampled windows only and should not be interpreted as a physiological scale, while actual training at each stage leverages all eligible samples within that stage. The visualization makes explicit three properties. First, engagement is staged and task-specific rather than relying on static monolithic usage of the full dataset. Second, a single MERDC-extracted novel cluster from the reservoir, denoted Task~6, initiates the incremental extension. Third, Tasks~1--5 are stably retained through replay without exhaustive reprocessing of their full raw distributions. Collectively, these behaviors indicate that the DIL + MERDC pipeline provides computation-aware, label-efficient, and distribution-selective adaptation for continual non-invasive glucose estimation under dynamic physiological streams.

	\subsection{Underlying PPG–Glucose Relationship and Model Interpretability}
	Blood glucose variation influences endothelial reactivity, smooth muscle tone, arterial stiffness and wave reflection, microvascular perfusion, hemorheology (viscosity, red cell deformability), and autonomic vasomotor modulation. These factors collectively shape PPG morphology across multiple temporal scales, encompassing rapid systolic upstroke dynamics, peak--shoulder--notch--dicrotic configuration and timing, late systolic and early diastolic contour shaped by reflection superposition, slow baseline and envelope drift, and inter-beat variability patterns \citep{c:3,c:5,c:8,c:9}. Hand-crafted approaches typically isolate a small subset of slope, interval, amplitude, or basic spectral descriptors, treating components as weakly coupled and limiting the representation of transitional segments such as the pre-peak shoulder, notch--post-notch interval, and late-diastolic low-slope region, as well as cross-scale context. A recurring concern is whether a data-driven model truly captures this multi-component physiological–optical relationship or merely relies on high-slope points \citep{c:10,c:11,c:16}. The present representation (Dynamic Incremental Learning, DIL) combines multiscale temporal encoding with adaptive updating to retain fast transient morphology, slower baseline modulation, and rhythm variability without collapsing them to a single dominant surrogate.
	
\begin{figure}[pos=t]
		\centering
		\includegraphics[width=0.90\linewidth]{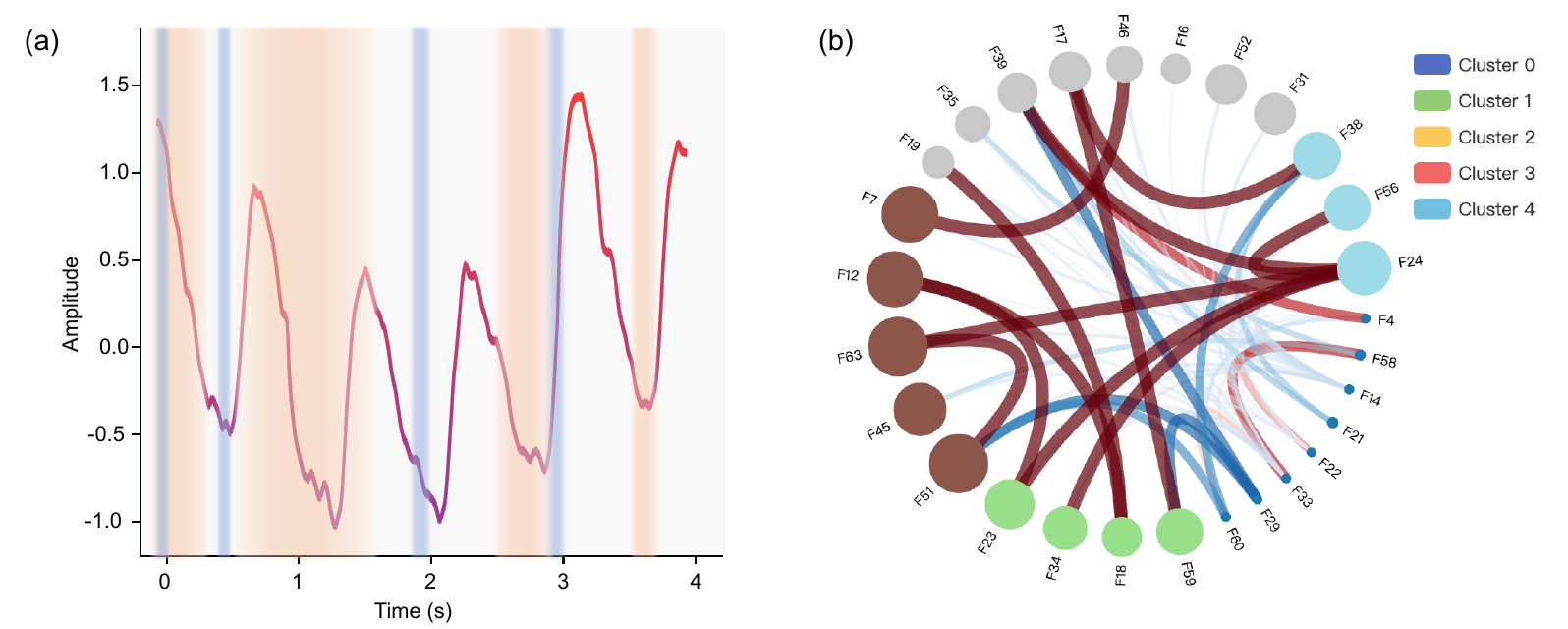}
		\caption{Comparison and characterization of predictive features. (a) Feature importance in PPG signals: derivative-based pseudo-gradients (blue) represent traditional hand-crafted features, while deep model Shapley attributions (orange) highlight distinct predictive time points. Values are interpolated for visualization clarity. (b) Chord diagram of DIL Backbone feature correlations in the test set: line thickness indicates correlation strength, color represents trend (red: positive, blue: negative), transparency denotes significance, and node size reflects feature importance. K-means clustering is applied for visual clarity.}
		\label{fig11}
		\vspace{-2em}
	\end{figure}

	Figure~\ref{fig11}(a) provides a direct comparative view: blue regions mark derivative-based pseudo-gradients corresponding to features emphasized by traditional slope- and amplitude-centric hand-crafted methods, while orange bands aggregate Shapley time-point attributions from the deep model \citep{c:64}.
	
	 Additional Shapley concentration on pre-peak shoulder inflections, the notch and immediate post-notch transition, and late-diastolic low-slope intervals indicates that predictive relevance extends beyond points of maximal first-derivative magnitude, reflecting dependence on composite morphology and contextual coupling rather than isolated steep segments, as in traditional slope- and amplitude-centric hand-crafted methods.
	
	Figure~\ref{fig11}(b) presents a chord diagram illustrating the correlation structure of high-dimensional features, which are derived from the feature vectors of our DIL backbone before mapping to blood glucose values and after pre-screening out features with variance less than 1e-6, across the entire test set, revealing intricate interdependencies among these features. This visualization underscores the complexity of the learned feature space, which captures nuanced relationships across temporal and morphological dimensions. The pattern aligns with a multifactor mapping from PPG to glucose, highlighting that the underlying relationship spans interacting temporal scales and morphological motifs that cannot be compressed into a small set of conventional descriptors. Together, the comparative attribution and latent structure indicate that the learned representation adaptively allocates relevance across physiologically coherent waveform components, rather than focusing solely on a narrow subset of hand-crafted features.
	\subsection{Model Robustness in Real-World Scenarios}
	Deploying beyond the lab introduces sensor imperfections, lifestyle dynamics, and environmental variability, and deep models are especially sensitive to signal quality \citep{c:9,c:53}. Our DIL framework reduces these risks without eliminating them entirely.
	
	\paragraph{Physiological and Lifestyle Variability.}
	Diet, activity, stress, and medication induce non-stationary intra-individual shifts. DIL adapts online to personalize representations, yet abrupt regime changes such as vigorous exercise or acute illness can temporarily degrade performance until sufficient data from the new state are incorporated.
	
	\paragraph{Signal Quality and Environmental Artifacts.}
	Real-world PPG is affected by motion artifacts, contact-pressure variation, temperature-dependent perfusion, and skin-tone differences. We keep preprocessing minimal to avoid overfitting to specific artifacts. Robustness mainly stems from the DIL backbone’s nonlinear modeling, which attenuates high-frequency noise while preserving physiological structure. In more demanding wearable studies, integrating real-time signal-quality assessment to flag low-confidence predictions and gate incremental updates would be an important next step, helping to prevent learning from irrecoverably noisy data.
	\subsection{Assumptions and Generalizability Limitations}
	A principal advantage of our proposed system is its capacity for continual learning, enabling it to adapt over time to the limitations discussed below.
	All models rely on assumptions, with generalizability limited by training data diversity. Our work is no exception.
	\paragraph{Dataset Representativeness.} Our longitudinal benchmark dataset, collected at a single center in China, may exhibit geographic and demographic homogeneity. PPG characteristics and physiological responses vary across ethnic groups due to genetic and lifestyle factors such as diet. Thus, model performance may not generalize to diverse populations. Future efforts should focus on creating large-scale, multi-center, ethnically diverse datasets for global validation.
	\paragraph{Reference Label Scope and Limitations.} The primary supervision in the main benchmark is reference capillary blood glucose measured using an FDA-cleared Accu-Chek Performa meter with daily L1/L2 quality-control verification, rather than laboratory venous plasma glucose. Although venous plasma assays remain the most stringent biochemical reference, standardized capillary SMBG acquired with a validated meter under routine quality control is an accepted and operationally appropriate reference strategy for longitudinal wearable-data collection \citep{c:27}. Accordingly, our benchmark should be interpreted as targeting agreement with high-quality point-of-care glucose references collected under realistic acquisition constraints, rather than direct equivalence to central-laboratory analytes. In addition, although PPG, reference glucose, and blood pressure measurements were completed within a two-minute acquisition window to preserve practical synchrony, residual asynchrony cannot be fully excluded when glucose-related vascular states evolve rapidly \citep{c:6,c:8,c:17}. This uncertainty does not invalidate the benchmark, but it should be borne in mind when interpreting absolute-error metrics and physiology-level correspondence.
	\paragraph{Aggregation Level and ISO-Related Interpretation.}
	The interpretation of window-level statistics also requires caution because each retained 40-s PPG acquisition is divided into ten non-overlapping 4-s windows, and these windows share the same reference capillary BG measurement. Adjacent windows therefore provide useful short-term waveform variability but should not be regarded as fully independent clinical measurements. To mitigate this issue, we reported ISO-related error-band results at the window, 40-s acquisition, and subject-macro levels. The acquisition-level analysis evaluates the averaged prediction from each 40-s acquisition, whereas the subject-macro analysis first computes subject-specific acquisition-level metrics and then averages them across subjects. Nevertheless, the ISO-related analysis in this study should be interpreted as a retrospective measurement-quality assessment using ISO 15197:2013 reference error bands, rather than as a formal regulatory compliance study. A dedicated prospective device-evaluation protocol with pre-specified sampling strata, independent clinical sites, and laboratory-grade reference measurements would be required before regulatory or deployment-oriented claims can be made.
	
	\paragraph{Underlying Physiological Assumptions.} The model assumes a stable, learnable link between PPG morphology and blood glucose levels, mediated by arterial stiffness and autonomic function \citep{c:3, c:8}. While plausible, this could be confounded by underrepresented comorbidities such as severe cardiovascular disease or advanced renal failure, or by medications affecting vascular tone. Further studies are needed to assess applicability in complex patient groups.
	\paragraph{Failure modes and safeguards of PGBA.} PGBA should be interpreted as a bounded mechanism for exploiting unlabeled PPG windows, not as a replacement for fully supervised learning. Its proxy gradients may become less reliable when PPG morphology is severely degraded by contact instability or low perfusion, when an unlabeled sample comes from a poorly represented physiological regime, when the sample lies in a sparsely observed high-glucose range, or when the reference glucose measurement contains non-negligible noise. To reduce the risk of biased updates, the present method uses bridge-loss training on labeled samples, proxy-gradient normalization to bound update magnitude, and replay-based gradient projection to prevent unlabeled updates from increasing previous-task memory losses. The reliability audit further checks whether proxy gradients are positively aligned with supervised gradients on labeled training-fold samples. Future work may investigate stricter confidence-based gating or consistency regularization, but these extensions are not required for the current results reported in this study.
	\paragraph{Future Work.} Future work will focus on three directions. First, larger multi-center and device-heterogeneous studies are needed to test whether the learned PPG--glucose relationship generalizes across broader demographic groups, skin conditions, sensor placements, and acquisition hardware. Second, prospective protocols with laboratory-grade reference glucose measurements and pre-specified sampling across hypo-, normo-, and hyperglycemic ranges will be necessary to move from retrospective benchmark analysis toward deployment-oriented validation. Third, the current DIL framework can be extended with complementary wearable modalities, such as ECG, accelerometry, skin temperature, or signal-quality gating, to improve reliability under motion, low perfusion, and sparsely represented high-glucose conditions. These directions are consistent with the present study's role as a proof-of-concept benchmark and method evaluation rather than a regulatory device-compliance study.
	\section{Conclusions}
	\label{sec:conclusions}
	This study proposes a dynamic incremental learning paradigm for non-invasive blood glucose estimation from PPG signals, combining three complementary components. First, a continual backbone with bounded episodic replay and gradient projection to balance stability and plasticity. Second, a mutual entropy-optimized replay-based dynamic clustering strategy to discover and maintain latent task structure under distribution shift, with ME2AC serving as its clustering core. Third, an uncertainty-quantified Proxy Gradient Bridging Agent (PGBA) to leverage unlabeled PPG windows in a label-efficient manner. We also establish a longitudinal benchmark with synchronized PPG, reference capillary blood glucose, and cuff blood pressure data from 183 participants over 285 days. In subject-independent 5-fold cross-validation, our approach achieved MAE \(0.64 \pm 0.01\) mmol/L and RMSE \(1.29 \pm 0.10\) mmol/L, with \(97.69 \pm 1.63\%\) of estimates in Clarke zones A+B. Additional ISO-related error-band analysis at the 40-s acquisition and subject-macro levels showed that the favorable error distribution was largely preserved beyond the original window-level evaluation, although these results should be interpreted as retrospective benchmark evidence rather than formal regulatory compliance. These findings establish a proof-of-concept for adaptive, label-efficient non-invasive glucose estimation under longitudinal drift.
	
	Notwithstanding these findings, limitations include the single-center cohort and fixed hardware, which may constrain generalizability across diverse populations, devices, and use contexts. Multi-center, device-heterogeneous, and prospective longitudinal evaluations are therefore warranted. Future work will prioritize integrating complementary modalities such as ECG and accelerometry with signal-quality gating for enhanced reliability, alongside developing lightweight, on-device continual learners supporting privacy-preserving personalization. Overall, our method provides a proof-of-concept for adaptive non-invasive glucose estimation in wearable sensing. We expect the controlled-access benchmark dataset and the detailed implementation protocol reported in this manuscript to support reproducible research and more rigorous benchmarking toward robust future wearable systems, while additional prospective validation and regulatory evaluation would still be required before translation beyond the research setting.
	
	\section*{Declaration of competing interest}
	The authors declare that they have no known competing financial interests or personal relationships that could have appeared to influence the work reported in this paper.

	\section*{Data availability}
	The benchmark dataset will be made available through a controlled-access repository upon submission of a completed access request form. Repository details are withheld for double-anonymized review and will be provided upon acceptance.
	
	\bibliographystyle{mymodel5}
	
	\bibliography{reference.bib}

\end{document}